\documentclass[11pt,a4paper]{article}                           

\usepackage{a4}          

\usepackage{amsmath}
\usepackage{amssymb}
\usepackage[usenames]{color} 

\usepackage{ifpdf}
\ifpdf
  \usepackage[pdftex,
    pdftitle={Massive Abelian Gauge Symmetries in F-theory},
    pdfauthor={Thomas W. Grimm, Max Kerstan, Eran Palti, Timo Weigand},
    pdfsubject={F-theory compactifications},
    bookmarksopen, bookmarksnumbered, bookmarksopenlevel=2]{hyperref}
\fi
\def\hybrid{\topmargin -20pt    \oddsidemargin 0pt
        \headheight 0pt \headsep 0pt
        \textwidth 6.25in       
        \textheight 9 in       
        \marginparwidth .875in
        \parskip 5pt plus 1pt 
          \jot = 1.5ex
   }
\hybrid
\numberwithin{equation}{section}
\numberwithin{table}{section}

\newcommand{\beq}{\begin{equation}}  \newcommand{\eeq}{\end{equation}}
\newcommand{\bal}{\begin{aligned}}   \newcommand{\eal}{\end{aligned}}
\newcommand{\bea}{\begin{eqnarray}}  \newcommand{\eea}{\end{eqnarray}}

\newcommand{\nn}{\nonumber}

\newcommand{\cO}{\mathcal{O}}
\newcommand{\cT}{\mathcal{T}}

\newcommand{\cP}{\mathcal{P}}
\newcommand{\cC}{\mathcal{C}}

\newcommand{\cL}{\mathcal{L}}
\newcommand{\cS}{\mathcal{S}}
\newcommand{\cK}{\mathcal{K}}
\newcommand{\cN}{\mathcal{N}}

\newcommand{\cA}{\mathcal{A}}

\newcommand{\cF}{\mathcal{F}}

\newcommand{\cV}{\mathcal{V}}

\newcommand{\tw}{\text{w}}

\newcommand{\I}{\text{Im}}
\newcommand{\R}{\text{Re}}

\newcommand{\be}{\begin{equation}}
\newcommand{\ee}{\end{equation}}

\newcommand{\half}{\frac{1}{2}}

\begin{document}

\baselineskip=14pt
\parskip 5pt plus 1pt

\vspace*{-1.5cm}
\begin{flushright}    
  {\small
MPP-2011-85\\
 CPHT-RR052.0711
  }
\end{flushright}

\vspace{2cm}
\begin{center}        
  {\LARGE Massive Abelian Gauge Symmetries and Fluxes in F-theory}
\end{center}

\vspace{0.75cm}
\begin{center}        
 Thomas W.~Grimm$^{1}$, Max Kerstan$^{2}$, Eran Palti$^{3}$, Timo Weigand$^{2}$
\end{center}

\vspace{0.15cm}
\begin{center}        
  \emph{$^{1}$ Max-Planck-Institut f\"ur Physik, 
               Munich, Germany } 
\\[0.15cm]
  \emph{$^{2}$ Institut f\"ur Theoretische Physik, Ruprecht-Karls-Universit\"at, \\
             Heidelberg, Germany}
             \\[0.15cm]
  \emph{$^{3}$ Centre de Physique Theorique, Ecole Polytechnique,\\ 
               CNRS, 
               Palaiseau, France
  }
\end{center}

\vspace{2cm}


\begin{abstract}

F-theory compactified on a Calabi-Yau fourfold naturally describes
non-Abelian gauge symmetries through the
singularity structure of the elliptic fibration. In contrast Abelian
symmetries are more difficult to study because of their inherently global
nature. We argue that in general F-theory compactifications there are
massive Abelian symmetries, such as the uplift of the Abelian part of the
$U(N)$ gauge group on D7-branes, that arise from non-K\"ahler resolutions
of the dual M-theory setup. The four-dimensional F-theory vacuum with
vanishing expectation values for the gauge fields corresponds to the
Calabi-Yau limit. We propose that fluxes that are turned on along these
$U(1)$s are uplifted to non-harmonic four-form fluxes. We derive the
effective four-dimensional gauged supergravity resulting from F-theory
compactifications in the presence of the Abelian gauge factors including
the effects of possible fluxes on the gauging, tadpoles and matter
spectrum.

\end{abstract}

\clearpage


\newpage

\tableofcontents

\section{Introduction}

Understanding Abelian gauge symmetries and their fluxes in F-theory \cite{Vafa:1996xn} (for recent reviews see \cite{Denef:2008wq})  is both of conceptual interest and of phenomenological importance.
F-theory has recently attracted revived interest from the perspective of string phenomenology due to its virtue of reconciling the idea of brane-localised gauge degrees of freedom with the appearance of exceptional gauge symmetry \cite{Donagi:2008ca}. In this context non-Abelian symmetries, whose geometric description in F-theory compactifications is well-understood, have been exploited rather heavily. An understanding of the geometric description of Abelian symmetries in F-theory has, however, remained rather elusive despite their being well under control as typical ingredients of perturbative Type II orientifold vacua with many applications to model building.
Not only in view of recent applications to model building but also from a more formal perspective it is therefore high time to expand on our knowledge of Abelian gauge symmetries in F-theory.

To appreciate the difference compared to $U(1)$ symmetries, recall that the non-Abelian gauge dynamics in F-theory are localised at the singularities of the elliptic Calabi-Yau fourfold $Y_4$. 
The structure of these degenerations has been under intense scrutiny since the early days of F-theory \cite{Morrison:1996na}. 
What makes a proper implementation of these degenerations technically involved is that the singularities really sit in the elliptic fiber. Consequently  a prerequisite for studying F-theory models including its phenomenological applications is a detailed understanding not only of the physical compactification space $B$, but of the fully fledged elliptic fibration over $B$. 
Motivated by the prospects of F-theory for GUT model building in the spirit of \cite{Donagi:2008ca}, a lot of recent effort has gone into the construction of F-theory compactifications to four dimensions, extending the technology of Calabi-Yau constructions with 7-branes from the Type IIB regime \cite{Blumenhagen:2008zz} into the non-perturbative one. 
 An example of an F-theory GUT compactification to four dimensions defined in terms of a  base space $B$  together with a Tate model thereon was found in \cite{Marsano:2009ym}.
The construction of fully-fledged elliptic fourfolds for F-theory compactifications with GUT physics was initiated in \cite{Blumenhagen:2009yv} using the toric framework developed for Calabi-Yau threefolds in \cite{Candelas:1997eh}.  In this approach the existence of the non-Abelian gauge dynamics can be guaranteed by an explicit resolution of the singular Calabi-Yau space. By now large classes of well-controlled fourfolds have been found \cite{Chen:2010ts}. 
The importance of the full resolution for a well-defined F-theory compactification with gauge dynamics and matter has more recently been stressed also in the studies \cite{Morrison:2011mb}.

Tracing back the geometric origin Abelian gauge bosons, on the other hand, is less immediate because these are in general not localised along singular divisors -  apart from the obvious exception of Cartan generators of a non-Abelian gauge group $G$. Nevertheless, an unambiguous detection of the presence of massless non-Cartan $U(1)$ gauge bosons hinges again upon a detailed understanding of the singularity structure of the fourfold and  in particular its resolution - this time, however, along complex codimension-two loci, i.e. curves. This was established in \cite{Grimm:2010ez} in the framework of the $U(1)$ restricted Tate model as a technically reliable method to guarantee a massless non-Cartan $U(1)$ symmetry in 4-dimensional F-theory vacua. For massless $U(1)$ bosons full control of the elliptic fibration is even more desperately needed than in the context of non-Abelian symmetries;  namely, as emphasized also in \cite{Hayashi:2010zp} $U(1)$s are sensitive to the full global details of a model.\footnote{In particular, while models without Abelian gauge bosons exhibit an encouraging match \cite{Blumenhagen:2009yv} between the Euler characteristic of the full resolved fourfold and that calculated by means of a semi-local spectral cover construction \cite{Hayashi:2008ba,Hayashi:2010zp}, this is no longer true in vacua with $U(1)$ symmetries \cite{Grimm:2010ez}.}
A different approach based on extending a $dP_9$ fibration over a 7-brane with non-Abelian gauge symmetry is advocated in \cite{Marsano:2010ix}.

Likewise, a proper understanding of gauge fluxes is expected to involve, at least for Abelian fluxes, a direct handle on the dual M-theory 4-form field strength $G_4$ defined globally on the fourfold, which so far has remained elusive.

In this article, our main focus is on the appearance of \emph{massive} Abelian gauge symmetries and their associated gauge fluxes in F-theory.
We propose to study these in the context of non-K\"ahler resolution of the singular fourfold $Y_4$ and give explicit formulae for their gauge fluxes in terms of 
a distinguished set of non-harmonic 2- and 4-forms.
 Our guiding principle is the structure of massive $U(1)$ symmetries in the perturbative regime of Type IIB orientifolds, whose F-theory uplift we investigate.
 While the massiveness of the $U(1)$ is encoded in a set of differential relations, many of the physically relevant quantities such as tadpoles and chirality turn out to involve only algebraic expressions. We expect these to generalise to massless $U(1)$s as well in a way that can be applied to global F-theory compactifications with $U(1)$ symmetries as considered in  \cite{Grimm:2010ez}.

Abelian gauge symmetries are ubiquitous in perturbative string theory. A subset of these that is especially important for phenomenological purposes comprises the $U(1)$ symmetries that arise from D-branes in type IIB orientifold compactifications. The canonical example is the diagonal $U(1)$ within the $U(N)$ gauge group associated to a stack of $N$ D7-branes. 
 This sector has a rich structure in IIB: the $U(1)$ symmetries can be even or odd under the orientifold action, they can have associated gauge fluxes turned on, they can be St\"uckelberg massive or not, anomalous or non-anomalous. 

As pointed out above, understanding this part of an F-theory compactification has essential bearings on phenomenological applications. In type IIB constructions the Abelian factors have been the central players in much of the resulting phenomenology affecting, as a brief sample: the chiral spectrum (see e.g. \cite{Blumenhagen:2006ci} for reviews), selection rules on field theory couplings \cite{Blumenhagen:2006xt}, moduli stabilisation and supersymmetry breaking \cite{Blumenhagen:2007sm}, mixing with visible sector  fields \cite{Burgess:2008ri}, gauge coupling unification \cite{Blumenhagen:2008aw} amongst many other features. In F-theory models $U(1)$ symmetries also have a central role, see \cite{Marsano:2009gv} and references therein for an incomplete list of uses. 

In contrast to the IIB case, the gauge group associated with a singularity of type $A_{N-1}$ in F-theory is  at first sight not $U(N)$, but $SU(N)$.
To see how the diagonal $U(1)$ can `disappear' as a massless gauge symmetry in uplifting from IIB to F-theory consider a IIB setup of a stack of D7-branes and an O7-plane wrapping 4-cycles in a CY three-fold which are not in the same homology class. In that case the diagonal $U(1)$ on the D7-stack always picks up a St\"uckelberg mass \cite{Ibanez:1998qp} by eating the orientifold odd axion arising from reducing the RR 2-form on the odd 2-form which is Poincar\'e dual to the odd combination of the D7 and O7 cycles \cite{Jockers:2004yj,Plauschinn:2008yd}. Since this $U(1)$ is always massive, a geometric F-theory uplift which only accounts for the massless $U(1)$s will completely miss it. As referred to above, the $U(1)$ can still play a crucial role in four-dimensional physics despite being massive, for example via flux turned on along it or by affecting the low energy selection rules left as a global symmetry. 

The key to understanding the nature of such $U(1)$s in F-theory is to consider the dual M-theory setup of a compactification to three dimensions. This 
allows to analyze the effective action and F-theory limit along the lines of \cite{Grimm:2010ks}. Then, as first proposed in \cite{Grimm:2010ez}, such $U(1)$s can be accounted for by considering non-K\"ahler deformations of M-theory where the $U(1)$ gauge-field arises from reducing the M-theory three-form $C_3$ on a non-closed two-from $\tw_0$. In this paper we build on this observation and complete it to a framework which can account for the structure associated to such $U(1)$s that is present in IIB. 

Our primary guiding principle is that the geometric structure we introduce in the M-theory setup should reduce, in the Type IIB limit, to well-known perturbative expressions involving D7-branes. We will incorporate type IIB quantities arising from compactifications involving D7-branes that are affected by the diagonal Abelian $U(1)$ factor in terms of a conjectured M-theory geometry. 

The particular IIB expressions we will reproduce are the four-dimensional supergravity and its gauging , the D3- and D5-tadpoles, and the chirality induced on intersecting D7-branes in the presence of gauge flux. Along the way we provide a detailed match of the 3-dimensional M-theory compactification uplifted to F-theory on the one hand and the Type IIB effective action with D7-branes on the other. 
Independently of this approach we will also provide complementary arguments for the conjectured geometry directly from studying the geometry of an $A_{N-1}$ singularity and by considering the uplift of the IIB base geometry to the elliptic fibration. 
The main challenge of this strategy of using the IIB limit as a guide is that more work is needed to see how our structure generalises to F-theory geometries with no IIB weak coupling limit such as those involving exceptional singularities. However, we believe our analysis is the natural starting point from which to attack such questions. 

The paper is composed as follows. In section \ref{gaugingIIB} we collect the relevant type IIB physics, including a careful assessment of all $\pi$'s and $i$'s,  which we aim to reproduce from F-theory. The following two sections form the core of our work.
In section \ref{sec:abeff}  we introduce a set of non-harmonic 2- and 4-forms on the elliptic fibration along with their intersection numbers in a way that will subsequently result in a complete match of the properties of $U(1)$ symmetries and gauge fluxes known from Type IIB limits. We express the $U(1)$ gauge fluxes in terms of these forms and demonstrate that the induced M5- and M2-tadpoles precisely match the D5- and D3-tadpole in Type IIB orientifolds. We also find a simple chirality formula that encapsulates the known expressions for the chiral index in the perturbative limit. 
Section \ref{sec:sugeff} is devoted to an in depth analysis of the effective action of M-theory reduction involving the set of non-harmonic forms introduced before.
We begin by laying out the 3-dimensional supergravity with special emphasis on the gauging induced by reduction of the M-theory 3-form along these  non-harmonic forms. Parts of this analysis are relegated to appendix \ref{dualization_appendix}. A detailed account of the F-theory lift then establishes a perfect match with the Type IIB gauging and D-terms in perturbative limits of F-theory.
In section \ref{sec:u1geometry} we discuss the geometry associated to the $U(1)$s thereby providing independent motivation for the structure proposed. We summarise our results in section \ref{sec:concl}.  

\section{Abelian Gauge Symmetries in Type IIB orientifolds}
\label{gaugingIIB}

We begin by reviewing the role played by the diagonal $U(1)$ arising on stacks of D7-branes in orientifold compactifications of Type IIB string theory. We will focus specifically on the gauging of Ramond-Ramond (R-R) scalars induced by the St\"uckelberg mechanism as well as the contribution to the D3- and D5-brane tadpoles from fluxes along such diagonal $U(1)$s. These aspects will turn out to be key in inferring the nature of the corresponding $U(1)$s and their fluxes in F-theory.

We consider a compactification of Type IIB string theory on  a Calabi-Yau threefold $X_3$, modded out by the orientifold action $\Omega (-1)^{F_L} \sigma$ \cite{Grimm:2004uq}. 
The holomorphic involution $\sigma$ acts on the K\"ahler and holomorphic three-form  $J$ and $\Omega$ as $\sigma^*J = J$, $\sigma^*\Omega = - \Omega$.
Under the induced action  of $\sigma$ the cohomology groups $H^{p,q}(X_3)$ split into $H^{p,q}_{\pm}(X_3)$.
Our notation for a basis of the 2- and 4-forms is summarized as follows: a basis of $H^{1,1}_+$ is denoted by $\omega_\alpha$ while its dual in $H^{2,2}_+$ is denoted by $\tilde \omega^\alpha$ with $\alpha= 1, \ldots h_+^{1,1}$. Moreover, the negative eigenspace $H^{1,1}_-$ has a basis $\omega_a$ and its dual $H^{2,2}_-$ a basis $\tilde \omega^a$ with $a=1, \ldots h_-^{1,1}$.
The two- and four-forms are dual in the sense that
\beq
\label{omega_omegatilde}
\int_{X_3} \omega_a \wedge \tilde \omega^{b} = \delta_a^b, \quad\quad \quad \int_{X_3} \omega_{\alpha} \wedge \tilde \omega^{\beta} =  \delta_{\alpha}^{\beta}. 
\eeq
Since the volume form is even under the involution the non-trivial intersection numbers are given by
\beq
\label{kappaOnCY}
\cK_{\alpha \beta \gamma} = \int _{X_3}  w_{\alpha} \wedge  w_{\beta} \wedge w_{\gamma} \in 2 \mathbb Z, \quad\quad \cK_{\alpha b c} = \int _{X_3}  w_{\alpha} \wedge  w_{b} \wedge w_{c} \in 2 \mathbb Z.
\eeq
It is important to note that due to the fact that we have chosen a basis with definite orientifold parity all triple intersection numbers will be \textit{even} integers, as given any intersection point the three divisors will also intersect at the orientifold image of this point.\\
In the Kaluza-Klein reduction the K\"ahler form $J$ of $X_3$ and the R-R and NS-NS forms enjoy an expansion 
\bea \label{JBC_expand}
&& J= v^{\alpha} \omega_{\alpha}, \quad\quad  C_2 = c^{a} \omega_a, \quad\quad B_2 \equiv   B_- + B_+ = b^{a} \omega_a + b^{\alpha} \omega_{\alpha}, \\
&& C_4 = c_{\alpha} \tilde \omega^{\alpha} + c_2^{\alpha} \wedge \omega_{\alpha} + c_4 + \ldots, \quad \quad
C_6 =  (\tilde c_2)_{a}  \wedge \tilde \omega^a + \dots. \nn
\eea
Here we have restrict ourselves to expansion along the even-dimensional cohomology as we will focus on Abelian gauge potentials arising from open strings propagating on D7-branes.\footnote{Expansion of $C_4$ along $H^3(X_3)$ leads to RR $U(1)$ fields. Furthermore we are not considering the $U(1)$ gauge factors from D3-branes at this stage.}
Note that in the above the component of the B-field $B_+$ along the even cycles, $b^{\alpha}$, is not a continuous modulus but can only take the discrete values $0, \frac{1}{2}$ consistent with the orientifold action.

\subsection{D7-branes and the St\"uckelberg coupling}
\label{sec:IIBStueckelberg}

Let us now consider the gauge dynamics of a stack of $N_A$ D7-branes along the holomorphic divisor $D_A$. 
The orientifold symmetry $\sigma$ maps $D_A$ to its orientifold image $D_A'  = \sigma^*  D_A$,
 so that in the upstairs geometry, i.e.~on the Calabi-Yau $X_3$ prior to orientifolding,  each brane is accompanied by
its image brane.  One distinguishes three qualitatively different classes of brane configurations:
\begin{enumerate}
\item \label{Dcase1} $[D_A] \neq [D_A'] \equiv [\sigma^* D_A]$,
\item \label{Dcase2} $[D_A] = [D_A']$ but $D_A \not= D_A'$ point-wise,
\item \label{Dcase3} $D_A = D_A'$ point-wise.
\end{enumerate}
Here the class $[D_A] \in H^{2}(X_3)$ is Poincar\'e dual to the divisor class $D_A$. We define the objects 
\beq
D_A^{\pm} = D_A \cup (\pm D'_A) \ ,
\eeq 
with Poincar\'e dual classes $[D^{\pm}_A] \in H^2(D_A)^{\pm}$. 
 Here $- D'_A$  is orientation reversed with respect to $D'_A$. Note that for orientifold invariant cycles one should include an extra factor of $\frac12$ to ensure $D_A^{+} =D_A$. 
This allows us to evaluate the corresponding wrapping numbers along the basis elements of $H_4^{\pm}(X_3, \mathbb Z)$ as
\beq
\label{wrappingnum}
C^{\alpha}_A =  \int_{D_A^{+}} \tilde \omega^{\alpha} \, , \qquad \quad 
C^{a}_A =  \int_{D_A^{-}} \tilde \omega^a  \, ,  
\eeq
 In the basis $(\tilde \omega^\alpha,\tilde \omega^a)$ normalized by \eqref{omega_omegatilde} the 
constants are actually integers characterizing the embedding of the D7-brane.
Note that in the last two cases 
one finds that $D_A^-$ is trivial in homology such that $C^a_A=0$.

Let us discuss which gauge theory arises from the above D7-brane configurations. The first situation gives rise to Abelian gauge bosons.
In the absence of gauge flux the  gauge group is $U(N_A)$ because upstairs $D_A$ and $D_{A}'$  each carry one $U(N_A)$ gauge factor, and the two are then identified under $\sigma$.
Note that a priori the gauge group is $U(N_A)=SU(N_A) \times U(1)_A$, not $SU(N_A)$, but the $U(1)_A$ is massive due to a St\"uckelberg mechanism \emph{even in the absence of gauge flux}, as will be reviewed in detail below. 
In the other two cases, the orientifold action projects the gauge group down to symplectic or special orthogonal gauge groups, depending on the details of the orientifold action. An important exception occurs for branes of type 2 that lie in the same homology class as the O7-plane, but are not placed on top of it.
Such configurations carry gauge group $U(N_A)$. The difference to branes of the first type is that in absence of flux the Abelian part does not become massive, as we will review below.

In the following we will concentrate on configurations leading to the gauge group $U(N_A)$, which contains a diagonal $U(1)$. The field strength $\hat{F}^A $ of such a stack $A$ appears in the Chern-Simons and DBI action only in the gauge
invariant combination ${ \mathbf F}^A_{D7}$ with the NS-NS B-field. The 
Cartan subsector of the brane gauge theory enjoys the expansion
\begin{equation}
2 \pi \alpha'\,  { \mathbf F}^A_{D7} = T_A^0 \, \Big(   2 \pi \alpha'\hat{F}_0^A  -  \i^* B_2 \Big) + 2 \pi \alpha'\sum_{i=1}^{N_A-1}  T^i_A\, \hat{F}_i^A.
\end{equation}
Here $T^0_A = 1_{N_A \times N_A}$ is the generator of the diagonal $U(1)_A \subset U(N_A)$, while $T_A^i$ are the generators of the Cartan of $SU(N_A)$.
We use the index range $I=\left(0,i\right)$ to label the full set of generators. 
The gauge field $\hat{F}_I^A $ splits into the field strength in four dimensions $F_{I}^{A}$ and the internal gauge flux $\cF_{I}^A$ along the cycle wrapped by the divisor
\beq
\hat{F}_{I}^A = F_{I}^A + \cF_{I}^A \;.
\eeq
The fluxes on the image stack $D_A'$ are given by $\cF_{I}^{A\prime}=-\sigma^*\cF_{I}^A$, where the minus sign is due to the worldsheet parity $\Omega$.
This $\cF^A_{I}$ will sometimes be further expanded as
\beq \label{cF_pm_expansion}
\cF^A_{I} =  \cF^{A,a}_{I} \omega_a  +  \cF^{A,\alpha}_{I} \omega_{\alpha} \;,
\eeq
where we are assuming, for notational simplicity, that the flux can be expressed in terms of the pull-back of two-forms from the bulk onto the brane and we are suppressing the explicit pull-back. Note that it is also possible to turn on flux which cannot be written as the pull-back of a cohomologically non-trivial bulk two-form. This type of flux will not be considered in the current analysis. 
Let us introduce some remaining pieces of notation.
It is convenient to combine the diagonal Abelian part of the gauge flux and the discrete background $B_+$-field as
\beq
\label{tildeF}
\tilde \cF^{A}_0 = 2 \pi \alpha' \cF_{0}^{A} -  \i^* B_+
\eeq
with components
\beq
\tilde \cF^{A,\alpha}_0 = 2 \pi \alpha' \cF_{0}^{A,\alpha} - b^{\alpha} \, ,\quad\quad \tilde \cF^{A,a}_0 = 2 \pi \alpha' \cF_{0}^{A,a}.
\label{frakfdef}
\eeq
The latter are the discrete combination of fluxes appearing, for example, in the chirality formulas summarized in 
appendix \ref{dualization_appendix} and are subject to the Freed-Witten quantisation condition \cite{Freed:1999vc}.

The diagonal part of the four-dimensional field strength, $F_{0}^A$, is special because it is well-known to acquire a mass via the St\"uckelberg mechanism. For D7-branes, there are actually 2 types of St\"uckelberg terms contributing to a mass of the Abelian gauge boson, see e.g. \cite{Plauschinn:2008yd,Grimm:2011dj} for recent expositions. In the sequel these will be referred to as the geometric and the flux induced couplings, respectively. A fact that will be crucial to matching the IIB configuration with F-theory is that the geometric mass terms depend only on the details of the compactification manifold  irrespective of the presence of gauge flux. Accordingly, the gauge boson of the diagonal $U(1)_A \subset U(N_A)$ \emph{always} acquires a mass term that cannot be switched off. The only exception is the case where $[D_A] = [D_A']$ but $D_A \neq D_A'$ pointwise. The analogue of  this mass term is automatically built in geometrically in F-theory. Both in Type IIB and in F-theory only special linear combinations of Abelian gauge bosons stemming from different branes can stay massless and give rise to a residual Abelian gauge symmetry. Therefore, in absence of gauge flux, Abelian gauge symmetry is non-local in nature since it depends on the interplay of several brane stacks. The  second, flux-induced type of mass terms, by contrast, are extra features that depend on the specific flux configuration. These terms yield an extra contribution to the full mass matrix.
Besides,  gauge flux which is not embedded into $U(N_A)$ diagonally can of course break the non-Abelian gauge groups and thus give rise to further  Cartan $U(1)$ symmetries which are not present in the fluxless case. 

The St\"uckelberg couplings are induced by interactions with the RR-background fields
\beq
\label{CS_couplings} 
 S_{CS}= - 2 \pi  \, \int_{{\mathbb  R}^{1,3} \times D_A} \sum_{p} C_{2p} \wedge   {\rm tr}\left[ e^{2 \pi \alpha'\,  { \mathbf F}^A_{D7}}\right]      \,  \sqrt{\frac{\hat A(TD)}{\hat A(ND)}}\, .
\eeq 
The terms relevant for the St\"uckelberg mechanism follow by dimensional reduction of this Chern-Simons coupling to linear order in $F^{A}_{0}$ taking into account both the contributions from the brane along $D_A$ and its orientifold image along $D_A'$. 
Due to the orientifold $\mathbb{Z}_2$ quotient one has to divide by 2 after adding the Chern-Simons actions of brane and image brane. Our conventions for the effective action are collected in appendix \ref{app:conventions}. The result contains two qualitatively different couplings
\beq
\label{St-coupl}
S_{St.} = - \frac{1}{4 \kappa_4^2 } \sum_A \Big(    Q_A^a \int_{{\mathbb R}^{1,3}}  F_{0}^A \wedge \big(\tilde c_{2\, a}  - \cK_{\alpha a c} b^c c_2^\alpha \big)  - Q_{A\alpha}  \int_{{\mathbb R}^{1,3}}  F_{0}^A \wedge  c_{2}^\alpha  \Big),
\eeq
where $(\tilde c_{2\, a} ,c_{2}^\alpha )$ are the two-forms appearing in \eqref{JBC_expand} which combine in the 
first integral to two-forms dual to the axionic scalars $c^a$. 
In this expression we have defined the constants
\bea
   Q_A^a &=&  2\pi\alpha' N_A C^a_A \;, \label{Qsodd} \\
   Q_{A \alpha} &=&  - 2\pi\alpha' N_A \Bigl( \cK_{\alpha \beta \gamma}  \, \tilde \cF^{A,\beta}_0 \, C^{\gamma}_A  +  \cK_{\alpha b c }  \, \tilde \cF^{A,b}_0   \, C^{c}_A \Bigr) ,\label{Qseven}   
\eea
where $\tilde \cF_A^{\alpha}$ and  $\tilde \cF_A^a$ were defined in \eqref{cF_pm_expansion} and (\ref{frakfdef}).\footnote{Note that turning on fluxes in the Cartan of the $SU(N_A)$ factor induces a coupling for the corresponding four-dimensional fields of the type (\ref{Qseven}). For gauge flux inherited from the bulk the commutant Abelian subgroup acquires a mass and contributes to the Fayet-Ilopoulos term. Both effects are absent only if the flux does not descend from two-forms defined on $X_3$.}
The first coupling $Q_A^a$ is completely geometric in that it only depends on the wrapping numbers and not on any fluxes. If these odd wrapping numbers are non-vanishing then the $U(1)_A$ will be massive as we will see in the next section. The second coupling $Q_{A\alpha}$ depends crucially on the fluxes $\tilde \cF_A^{\alpha}$ and  $\tilde \cF_A^a$ such that if all the fluxes are turned off this does not contribute to the $U(1)$ mass.

\subsection{The effective four-dimensional action}
\label{sec:IIBsugeff}

Let us now discuss the couplings  \eqref{St-coupl} as part of the four-dimensional effective action 
obtained by reducing the orientifold set-up with D7-branes.
Our conventions for the canonical $\cN=1$ supergravity form for the four-dimensional effective action are  \cite{Wess:1992cp}
\begin{multline} \label{eq:N_1}
   \mathcal{S}^{(4)}_{\mathcal{N}=1}=\frac{1}{\kappa_4^2}
      \int \Big[ -\tfrac12 R\:*_41 - \:K_{M\bar N}\nabla M^M\wedge*_4\nabla\bar M^{\bar N} \\
      - \tfrac12 \mathrm{Re} f_{AB}\, F^A\wedge*_4F^B
      - \tfrac12 \mathrm{Im} f_{AB}\, F^A\wedge F^B
      -\left(V_{\rm F}+V_{\rm D}\right)\:*_41 \Big] \ ,
\end{multline}
where
\beq \label{eq:V}
   V_{\rm F}=e^K\left(K^{M\bar N}D_MW
      D_{\bar N}\bar W-3\left|W\right|^2\right) \ , \qquad \quad 
   V_{\rm D}=\tfrac{1}{2}\left(\mathrm{Re} f\right)^{-1\;AB}D_A D_B \ .
\eeq

The couplings determined in equation \eqref{St-coupl} result in a gauging of the shift symmetry of the axions $c^a$ and $c_\alpha$ after their respective dual two-forms $\tilde c_{2\, a}$ and $c_{2}^\alpha$ are eliminated from the four-dimensional effective action. The four-dimensional chiral fields containing these axions are given by\footnote{Note that the definitions of the complex coordinates $\tau$ and $T_\alpha$ are corrected when including 
massless fields arising from the open string sector of the D7-branes. In particular, 
the D7-brane deformations correct $\tau$, while the $T_\alpha$ are modified in the presence 
of Wilson line moduli.} 
\bea
G^a &=& c^{a} - \tau b^a, \nn \\
\label{def_Talpha}
T_{\alpha} &=& \frac{1}{2}  \cK_{\alpha \beta \gamma} v^\beta v^\gamma + i \left(c_{\alpha} - \frac12  \cK_{\alpha bc} c^b {b}^c \right)  + \frac{i}{2\left(\tau-\bar{\tau}\right)} \cK_{\alpha bc} G^b \left( G^{c} - \overline{G}^c\right) \label{t_alpha}  \nn \\
&=&  \frac{1}{2}\cK_{\alpha \beta \gamma} v^\beta v^\gamma+ i \left(c_{\alpha} - \cK_{\alpha bc} c^b {b}^c \right) + \frac{i}{2} \tau \cK_{\alpha b c}  b^b   \, b^c \;,
\eea
where $\tau = C_0 + i\, e^{-\phi}$ represents the axio-dilaton.
The gauging of the axions results in the appearance of covariant derivatives for these chiral fields of the form
\bea
   \nabla G^a &=& d G^a - Q_A^a A^A,  \label{gauging1} \\
   \nabla T_\alpha &=& dT_\alpha - i Q_{A \alpha} A^A . \label{gauging2}
\eea 
This implies that the $U(1)_A$ are non-linearly realized with field-independent $Q_{A}^a,\ Q_{A\alpha}$. It is thus straightforward to determine the corresponding Killing vectors 
\beq \label{Killing_sum}
   X_A^a = - Q_A^a \ , \qquad  X_{A \alpha} = - i Q_{A \alpha}\ .
\eeq 

For later purposes we note that the gauge kinetic function of the D7-brane stack is given in terms of the chiral fields as
\be
\label{gaugeKinIIB}
f_{AA} = \frac{1}{4} (2\pi\alpha')^2 N_A C^\alpha_A T_\alpha,
\ee
where we neglect flux-induced contributions not relevant to our present analysis. 
Using the fact that the effective theory is $\cN=1$ supersymmetric one readily evaluates 
the D-terms induced by the gaugings \eqref{gauging1} and \eqref{gauging2}. 
This requires the K\"ahler metric for the chiral multiplets.
Explicitly, the kinetic terms of the moduli\footnote{In addition to $G^a$ and $T_\alpha$ these include the complex structure moduli as well as the axio-dilaton $\tau$.} are encoded in the K\"ahler potential 
\beq \label{KpotIIB}
   K = - \log \Big[-i \int_{X_3} \Omega \wedge \bar \Omega \Big] - \log\big[-i(\tau - \bar \tau)\big] - 2\log[\cV]\ ,
\eeq
where $\cV=\frac{1}{3!}\int_{X_3} J \wedge J\wedge J$ is the volume of $X_3$ evaluated in the 
ten-dimensional Einstein frame.
The 
general expression for the D-terms is 
\beq \label{genD}
     i \partial_{I} D_A = K_{I \bar J} \bar X^J_A\ , 
\eeq
where $X_A^J$ are the Killing vectors of the gauged isometry, and $K_{I\bar J}$ is the 
K\"ahler metric as determined from the K\"ahler potential \eqref{KpotIIB}. Let us consider 
the case without gauged matter fields.
Using the fact that the Killing vectors \eqref{Killing_sum} are constant, one integrates
\eqref{genD} to $ K_{\bar J} \bar X^J_A = i D_A$. The first derivatives of the 
K\"ahler potential \eqref{KpotIIB} are evaluated to be 
\beq
   \partial_{G^a} K = - \frac{i}{2 \cV } \int_{X_3} \omega_a \wedge J \wedge B  \ , \qquad \partial_{T_\alpha} K = -\frac{v^\alpha}{2 \cV} .
\eeq
One thus concludes that the $U(1)_A$ D-term is given by 
\bea
\label{DtermIIB}
 \frac{1}{2\pi\alpha'} D_A &=&  \frac{N_A}{2 \cV} \int_{D_A} J \wedge ( 2 \pi \alpha' \cF_A - \i^* B_2 )\\
          &=&  \frac{v^\alpha}{2 \cV} N_A  \big(    \cK_{\alpha \beta \gamma} \tilde \cF^{A,\beta}_0 C_A^\gamma  +     \cK_{\alpha ac} \tilde \cF^{A,c}_0 C_A^a   - \cK_{\alpha ac} b^c C_A^a   \big)\ .
 \nn
\eea	
Here the term involving $b^a$ arises from the gauging of $G^a$ and is universally present even in 
case all fluxes are set to zero. The remaining terms arise from the gauging of $T_\alpha$
and are absent if the discrete data corresponding to the gauge fluxes and the discrete B-field are zero.

The mass term for the diagonal $U(1)$ induced by the St\"uckelberg gauging takes the general form $m^2_{AB} \propto  K_{I \bar J} X^I_A \bar X^J_B$. To obtain the physical mass it is necessary to diagonalize the kinetic terms and then rescale the gauge fields to bring the kinetic terms into canonical form. To illustrate the nature of the purely geometric mass term let us for simplicity focus only on a single stack of D7-branes and set the gauge fluxes $\tilde \cF$ to zero. Using the gauge kinetic function \eqref{gaugeKinIIB} to rescale the gauge fields one obtains the mass\footnote{The K\"ahler metric is given in \cite{Grimm:2004uq}.}
\bea
\label{massIIB}
m^2 &=& -\frac{2}{\R f_{AA}} K_{G^a \bar G^b} N_A^2 C^a_A C^b_A  \nn \\
&=& 4 N_A C^a_A C^b_A  \Big( C^\alpha_A \cK_\alpha - e^\phi C^\alpha_A \cK_{\alpha b c} b^b b^c \Big)^{-1} \nn \\
&& \times \Big[ \frac{e^\phi}{\cV} \cK_{a b} -\frac{1}{2\cV^2} \cK_{ac}\cK_{bd} b^c b^d +\frac{1}{\cV} \cK^{\alpha \beta}\cK_{\alpha a c}\cK_{\beta b d} b^c b^d \Big].
\eea
Note that the first, purely geometric, term in the square brackets roughly scales with the volume as $\cV^{-\frac{2}{3}}$ while the terms dependent on the moduli $b^a$ scale as $\cV^{-\frac{4}{3}}$ and are therefore sub-leading contributions in a large volume expansion. The geometric St\"uckelberg mass is therefore suppressed by a factor of $g_s$ with respect to the  Kaluza-Klein scale.\footnote{In anisotropic compactifications, due to the different Kaluza-Klein scales associated with brane and bulk cycles the precise value of the St\"uckelberg mass can vary drastically for anomalous versus non-anomalous $U(1)$s. See e.g. \cite{Conlon:2008wa} for details.}

Finally, let us consider brane stacks of the type $[D_A] = [D_A']$ that lie in the same homology class as the O7-plane but do not coincide with it.
In this case, the gauge is $U(N_A)$, and in absence of flux the $U(1)_A$ factor survives as a massless gauge symmetry. The reason is that the geometric St\"uckelberg terms rely on non-zero wrapping numbers $C^{a}_A$ with respect to involution-odd classes, which vanish for branes in the same class as the O7-plane. Consistently, there arises no $B_-$ dependent Fayet-Iliopoulos D-term since the $B_-$-field does not couple to $D_A$.

\subsection{The induced D5- and D3-tadpoles in IIB}
\label{sec:d5tadiib}

As is well known the Chern-Simons couplings of $C_8$ and $C_6$ in \eqref{CS_couplings} give rise to modified Bianchi identities for the respective dual field strengths $F_1=d C_0$ and $F_3 = d C_2$. 
The equation of motion for $C_8$, once we include both the branes and images as well as the orientifold contribution and impose the suitable duality relation, gives\footnote{Note that to derive the tadpole constraints in the democratic formulation one needs to take into account an extra factor of $\tfrac{1}{2}$ in the Chern-Simons action of D-branes and O-planes in order to capture only the electric couplings. This is consistent with the derivation of the D3-tadpole via the Bianchi identity of the self-dual $F_5$ as  e.g. in \cite{Giddings:2001yu}.}
\be
dF_1 = \frac12 \sum_A N_A\, \delta( D_A^+ ) - 4 \, \delta(D_{O7}) \;.
\ee
The $\delta(D)$ are delta-currents localizing on the divisor $D$, i.e.~sharply
 localized two-forms on the D7-branes $D_A^+$ and the orientifold plane $D_{O7}$.  These are given by $\delta(D_A^+) = C^\alpha_A \omega_\alpha$, $\delta(D_A^-) = C^a_A \omega_a$ in terms of the wrapping numbers $(\ref{wrappingnum})$. The factor of $\tfrac12$ on the right hand side of this relation is a consequence of the orientifold geometry. It is consistent with the fact that in order to evaluate the monodromy of, say, a probe D(-1)-brane one must consider loops which encircle both the brane and the image brane, resulting in an integer monodromy \cite{Collinucci:2008pf}.

Consistency clearly requires the right hand side to vanish in cohomology imposing the 
D7-brane tadpole cancellation condition 
\beq
\sum_A  N_A\, ([D_A] + [D_A']) = 8 [D_{O7}] \ ,
\eeq
which has to be evaluated in cohomology, e.g.~by 
integrating over a basis of $H_2(X_3)$.

Similarly one can proceed to evaluate the D5-brane tadpole constraint. The equation of motion of $C_6$ leads to
\be
dF_3 = - \frac12  \sum_A N_A\, \left[ (2 \pi \alpha' {\cal F}^{A,-}_0 - \i^* B_2^-) \wedge \delta(D_A^+ ) +  ( 2 \pi \alpha' {\cal F}_0^{A,+} - \i^* B_2^+) \wedge  \delta(D_A^- )  \right] + 4 \i^* B_2 \wedge \delta(D_{O7} ) \;, 
\ee
where we have defined ${\cal F}^{A,\pm}_0 = \half \left({\cal F}_0^A \pm \sigma^* {\cal F}_0^A \right)$ and similarly for $B_2^\pm$.
We see that if we turn on flux along the diagonal $U(1)$ of the D7-brane stack we induce a D5-tadpole
\be
dF_3= - \frac12 \sum_A N_A\, \left[ \tilde \cF_0^{A, \alpha} \omega_{\alpha} \wedge \delta(D_A^-)+( \tilde  \cF_0^{A, a} - b^a)\omega_{a} \wedge\delta(D_A^+)] \right] + 4 b^a \omega_a \wedge \delta(D_{O7})\; , \label{d5tad1}
\ee
with constant fluxes $\tilde \cF_A^{\alpha}$,  $\tilde \cF_A^a$ defined in \eqref{cF_pm_expansion} and (\ref{frakfdef}).
Note that the $b^{a}$ moduli appear in the expression (\ref{d5tad1}) though if we integrate it over the basis of $H_4(X_3)$ to extract the tadpoles we can use the D7-tadpoles to eliminate this dependence on the continuous modulus \cite{Blumenhagen:2008zz,Plauschinn:2008yd} to yield the 
integrated D5-tadpole constraint
\beq
\label{D5tadInt}
  \frac12 \sum_A N_A \big( \cK_{\alpha b c} \tilde  \cF_0^{A, \alpha} C^b_A+ \cK_{a \beta c} \tilde  \cF_0^{A,a}  C^\beta_A  \big)= 0\ .
\eeq

The flux also induces a D3-tadpole which takes the form 
\bea
N_{D3} + N_{\rm gauge} = \frac{N_{03}}{4} + \sum_A N_A \frac{\chi_0(D_A)}{24} + \frac{\chi(D_{07})}{6},
\eea
with $N_{D3}$ and $N_{O3}$ counting the number of D3-branes and O3-planes,  while $\chi(D_{07})$ and $\chi_0(D_A)$ denote the Euler characteristic of the O7-plane and the modified (in the sense of \cite{Collinucci:2008pf}) Euler characteristic of the 7-branes. We focus here on
the gauge flux contribution due to the diagonal $U(1)_A$ flux given by \footnote{For a treatment of the Cartan fluxes of $SU(N_A)$ see e.g.  \cite{Blumenhagen:2008zz}.}
\be
\begin{array}{rl}
N_{\rm gauge}^{(0)} =& - \frac{1}{4}  \, \sum_A \, N_A \Big( \int_{D_A}   \tilde {\cal F}^A_0 \wedge \tilde {\cal F}^A_0  + \int_{D_A'} \tilde {\cal F'}^A_0 \wedge \tilde {\cal F'}^A_0 \Big)  \\ \\
        =& - \frac{1}{4} \sum_A N_A \Big( \cK_{\alpha \beta \gamma} C^\alpha_A \tilde \cF_0^{A, \beta} \tilde \cF_0^{A, \gamma} +   \cK_{\alpha b c} C^\alpha_A  \tilde \cF_0^{A,b}  \tilde \cF_0^{A, c} + 2    \cK_{a b \gamma} C^a_A \tilde \cF_0^{A, b}  \tilde \cF_0^{A, \gamma}             \Big). \label{d3iibNgauge}
\end{array}
\ee

In section \ref{sec:d5tad} we will see how fluxes along the (massive) diagonal $U(1)$s and the resulting tadpoles \eqref{d5tad1} are uplifted to F-theory.

\section{Massive $U(1)$ symmetries  and their fluxes in F-theory}
\label{sec:abeff}

In this section we formulate our proposal for describing massive Abelian symmetries in F-theory. 
The key aspects of the Type IIB compactifications reviewed in section \ref{gaugingIIB} 
will be our guide in formulating the F-theory setup. In section \ref{sec:massless_reduction} 
we begin with the well-known implementation of the Cartan $U(1)$s that lie within the non-Abelian 
part of the gauge group as resolution cycles in M-theory. In section \ref{nonKaehlergeometry} 
we introduce the details of our proposal, namely we present the 
 general form of the non-K\"ahler deformations in M-theory which give rise to 
the massive $U(1)$s in F-theory.  
We have three independent pieces of evidence for this picture:
In section \ref{sec:d5tad} we give a proposal for the form of the 
M-theory uplift of orientifold even and odd IIB fluxes and check its validity 
 by uplifting the IIB D3- and D5-tadpoles.
In section \ref{sec:iibfluxchi} we present a formula for the chiral index induced by these $G_4$-fluxes directly in F-theory
which is in agreement with the known analogue for Type IIB fluxes.
The last piece of evidence will then be presented in section \ref{sec:sugeff}, where we compute the resulting F-theory effective 
action, finding complete match with the Type IIB expressions of section \ref{gaugingIIB}.

\subsection{Cartan U(1)s from resolution of divisors} 
\label{sec:massless_reduction}

Let us consider F-theory compactified on an elliptically fibered 
Calabi-Yau fourfold $Y_4$. The complex structure of the 
two-torus fiber of $Y_4$ corresponds to the axio-dilaton 
$\tau$ varying over the base $B_3$.
We denote the projection to the base by
\beq
\label{pi}
\pi: Y_4 \longrightarrow B_3,
\eeq
with 
7-branes on divisors $D^{\rm b}_A$ in the base $B_3$. The location 
of these 7-branes is identified by locating the singularities 
of the elliptic fiber. These are encoded by the discriminant 
$\Delta$. In general, the 
7-branes can admit non-Abelian gauge-groups $G_A$, which implies 
that $Y_4$ is singular itself. 
While the Tate algorithm for singular 
elliptic fibrations often provides an algorithm to read off the 
non-Abelian gauge groups of an F-theory compactification, determining the appearance of 
Abelian vector multiplets must be addressed separately.

The four-dimensional $\cN=1$ effective action of F-theory compactified on an 
elliptically fibered Calabi-Yau fourfold $Y_4$ 
has been studied via an M-theory lift in \cite{Grimm:2010ks}. 
Taking the detour via M-theory is necessary 
since there is no twelve-dimensional effective action for F-theory. 
Instead one lifts the three-dimensional effective theory obtained by 
compactifying M-theory on $Y_4$ by performing an appropriate scaling 
limit. The basic idea is 
to fiberwise apply the duality between M-theory on $T^2$ and Type IIB string theory on $S^1$. 
More precisely, one reduces M-theory on one of the two one-cycles of $T^2$ to obtain type IIA 
string theory with generically varying dilaton. T-duality along the second one-cycle of $T^2$
leads to the corresponding Type IIB set-up with varying dilaton. Fibering the $T^2$ over some base 
$B_3$ leads to Type IIB string theory on $B_3 \times S^1$. Due to the T-duality operation, 
the $S^1$ is decompactified in the limit where the size $R$ of the original $T^2$ goes to zero.
Hence, one of the dimensions of $Y_4$ becomes a non-compact space-time dimension and one obtains 
a four-dimensional set-up from the M-theory compactification on $Y_4$.
This is known as the F-theory limit.

In case the 7-branes carry non-Abelian gauge-groups $G_A$
the fourfold $Y_4$ admits singularities. In this case one has to resolve 
the singularities to obtain a smooth Calabi-Yau fourfold $\hat Y_4$
in order to define the topological data such as intersection 
numbers and Chern classes.
In performing the resolution 
one introduces two-forms 
\beq \label{twset}
   \tw_{iA} \ \ \in\ \ H^2(\hat Y_4,\mathbb{Z}) \ ,  \qquad i = 1, \ldots , \text{rank}(G_A)\ , \quad A = 1,\ldots, n_7\ ,
\eeq
where $A$  labels one of the $n_7$ 7-brane stacks under consideration. 
In general, the resolution of singularities is a hard task. However, a well-known and large class 
of examples is provided by studying hypersurfaces or complete intersections
in a toric ambient space. In these cases the resolution can be performed 
explicitly, and cohomologically non-trivial $\tw_{iA}$ are constructed \cite{Candelas:1997eh,Blumenhagen:2009yv,Chen:2010ts}.

Let us briefly recall how the two-forms $\tw_{iA}$ appear in the dimensional reduction 
of M-theory to three space-time dimensions. Note that the $\tw_{iA}$ are not the 
only elements in $ H^2(\hat Y_4,\mathbb{Z})$. In addition one finds 
$h^{1,1}(B_3)+1$ elements which arise as follows. There is one two-form $\omega_0$
with indices along the elliptic fiber which is Poincar\'e dual to the base $B_3$.
In addition there are $h^{1,1}(B_3)$ two-forms $\omega_\alpha$ 
which are Poincar\'e dual to non-trivial divisors $D_\alpha = \pi^{-1}(D_\alpha^{\rm b})$, where $D_\alpha^{\rm b}$ is 
a divisor of $B$ and  $\pi: Y_4 \rightarrow B$ is the projection to the base as introduced in \eqref{pi}.
For simplicity, let us not perform a general dimensional reduction and assume that there 
are no massless fields arising from the reduction in three-forms on $\hat Y_4$ by 
considering examples with  $h^{2,1}(\hat Y_4)= 0$, i.e.~with a trivial third 
cohomology $H^{2,1}(\hat Y_4)$.\footnote{For a non-trivial $H^{2,1}(\hat Y_4)$, one finds 
$h^{2,1}(B_3)$ vector fields in the F-theory lift, and $h^{2,1}(\hat Y_4)-h^{2,1}(B_3)$ 
complex scalars in the F-theory lift. The $h^{2,1}(B_3)$ vectors are R-R 
bulk $U(1)$ gauge fields in F-theory.}
More precisely, we consider the expansion
\beq \label{C3_expand_simple}
  C_3 = A^0 \wedge \omega_0 + A^\alpha \wedge \omega_\alpha + A^{i A} \wedge  \tw_{i\, A} 
        \ ,
\eeq
where $(A^0,A^\alpha , A^{iA})$ are vectors in three dimensions. 
Clearly, in a three-dimensional effective theory with ${\cal N}=2$ supersymmetry corresponding to ${\cal N}=1$ in four dimensions
the vectors $(A^0,A^\alpha , A^{iA})$  are accompanied by real scalars $(v^0,v^{\alpha},  v^{iA})$ in their supermultiplets.
These scalars arise in the expansion of the K\"ahler form $J$ on $\hat Y_4$ as
\beq \label{J_expand_simple}
   J =  v^0 \omega_0  + v^\alpha \omega_\alpha + v^{i\, A} \tw_{i\, A}    \ .
\eeq
In the F-theory lift the three-dimensional vector multiplets $(A^0,v^0)$, $(A^\alpha,v^\alpha)$ and 
$(A^{iA},v^{iA})$ are identified as follows \cite{Grimm:2010ks}: 
\vspace*{-.3cm}
\begin{itemize}
\item \textit{Geometry:}  $(A^0,v^0)$ will become part of the four-dimensional space-time metric. 
The circumference $r$ of the fourth dimension is related to  $v^0$ as 
\beq  \label{def-R}
        R \equiv \frac{v^0}{\cV} = \frac{1}{ r^2}\ ,
\eeq 
where $\cV$ is the volume of the Calabi-Yau fourfold $\hat Y_4$.
$r$ becomes infinite in the F-theory limit $R \rightarrow 0$.
The vector $A^0$ is the off-diagonal component of the metric in the compact fourth direction. 

\item \textit{Chiral Multiplets:} $(A^\alpha,v^\alpha)$ become four-dimensional chiral multiplets, the complexified 
K\"ahler moduli $T_\alpha$ of the base $B_3$. In the F-theory limit finite volumes are parametrized by the combination 
\beq
   L^\alpha \equiv \frac{v^\alpha }{ \cV}\ .
\eeq

\item \textit{Vector Multiplets:} $(  A^{iA}, v^{iA} )$ become four-dimensional vectors, corresponding to the 
Cartan $U(1)$ gauge bosons of the 7-branes.  Note that for an ADE gauge group $G_A$ there are ${\rm rank}(G_A)$ such $U(1)$ gauge bosons. The relevant 
field in the F-theory limit is
\beq
   \xi^{iA} \equiv \frac{v^{iA}}{\cV} =  \frac{\zeta^{iA}}{r^2}\ ,
\eeq  
where $\zeta^{iA}$ is the actual fourth component of the four-dimensional Cartan $U(1)$ gauge bosons. 

\end{itemize}

The F-theory limit is defined by setting the background values around which the fields 
$R,L^\alpha$ and $\xi^{iA}$ have to be expanded. While $L^\alpha$ are finite in the 
F-theory limit and parameterize the volumes in the base $B_3$, the fields $R,\xi^{iA}$
go to zero in the vacuum. We assign the scaling behavior
\beq \label{F-limit_scalings}
   R\ \propto\ \epsilon^{3/2}\ ,\quad  \qquad \xi^{iA}\ \propto \ \epsilon^2\ ,
\eeq 
to perform this limit $\epsilon \rightarrow 0$. This allows 
to distinguish terms which have to be kept from those which are dropped in 
the F-theory limit decompactifying to four dimensions \cite{Grimm:2010ks}. Let us
stress that in M-theory all fields are physical, and the fluctuations of the 
scalars $R,\xi^{iA}$ around the discussed limit have to be kept in the 
spectrum to perform the three- to four-dimensional lift.

The vector multiplets $(A^{iA},v^{iA})$ 
only capture the degrees of freedom in the Cartan subalgebras of $G_A$.
Suppose the elliptic fourfold $Y_4$ acquires a singularity of type $G_A$ over the divisor $D^{\rm b}_A$ in the base $B_3$. The pullback of $D^{\rm b}_A$ to the resolved space $\hat Y_4$ will be denoted as $D_A = \pi^{-1} (D^{\rm b}_A)$ with associated class
\bea
\label{DA}
[D_A] = C_A^\alpha \, \omega_\alpha.
\eea
Then the group theory of $G_A$ along $D_A$ is encoded in the intersection numbers
\bea
 \label{intersection-resolution}
   \int_{\hat Y_4} \omega_\alpha \wedge \omega_\beta \wedge \tw_{iA} \wedge \tw_{jB}  = - \delta_{AB}\, \cC^B_{ij}\, C^\gamma_A \int_{B_3} \omega_\alpha \wedge \omega_\beta \wedge \omega_\gamma \;.
\eea
Here $\cC^A_{ij}$ is the Cartan matrix of the gauge group $G_A$. 
To fully enhance the 
gauge group to $G_A$ one has to also include M2-branes becoming massless in the blow-down limit 
$\hat Y_4 \rightarrow Y_4$. 

Following the classification above, the number of massless Abelian gauge factors 
present on the original, singular fourfold $Y_4$ with non-Abelian gauge groups $\prod_A G_A$ is given by
\beq
\label{nU(1)}
 n_{U(1)} = \Big( h^{1,1}(\hat Y_4) - h^{1,1}(B_3) - 1 - \sum_A \text{rank}(G_A) \Big) + h^{2,1}(B_3)\ ,
\eeq
where the expression in the brackets counts the number of $U(1)$ factors on the non-Abelian 7-branes.
One additionally has $h^{2,1}(B_3)$ vectors corresponding to the expansion of the R-R four-form $C_4$.

\subsection{Non-harmonic forms for massive $U(1)$s} 
\label{nonKaehlergeometry}

In the previous section we described $U(1)$s that are the Cartan elements of the 
non-Abelian gauge groups present in F-theory compactifications. 
These do therefore not account for the additional $U(1)$s present in type IIB orientifolds as 
the diagonal Abelian factor in the $U(N)$ gauge 
group of a stack of $N$ D7-branes. In this section we will introduce a new class 
of F-theory $U(1)$ gauge fields which we claim correspond to the uplift of the 
additional IIB $U(1)$s. The key idea we wish to implement was presented originally 
in \cite{Grimm:2010ez}. We start by recalling this proposal and then elucidate and expand on it. 

As reviewed in section \ref{gaugingIIB}, in type IIB orientifolds if a brane stack and its orientifold image are not cohomologous the diagonal $U(1)$ factor becomes massive through a St\"uckelberg mechanism. Such a mass is geometric in nature in that its presence is independent of any fluxes. Therefore we expect that its mass in F-theory must also be geometric even in the absence of flux. There is a well-known geometric mechanism to give masses to scalar moduli and closed-string gauge fields in type IIB supergravities by allowing for non-harmonic forms in the dimensional reduction \cite{Gurrieri:2002wz}. In \cite{Grimm:2010ez} it was suggested that a similar mechanism is at work for open-string gauge fields. Here the non-harmonic structure is not associated just to the base manifold but also to the elliptic fibration. 

As is evident from the scaling of the mass formula for the $U(1)$s  massive through the 
gauging \eqref{gauging1} in Type IIB,  the mass of these $U(1)$s is at the Kaluza-Klein 
scale in F-theory. In other words there is no parametric separation between the $U(1)$ mass 
and the other Kaluza-Klein modes. However, as we will show below, it is still possible to make the massive $U(1)$ visible 
by including certain non-harmonic forms in the dimensional reduction. 
These non-harmonic forms are present already in the Calabi-Yau fourfold $\hat Y_4$. 
In the three-dimensional M-theory reduction on  $\hat Y_4$, the massive $U(1)$s derive from the fluctuations of  $C_3$ and the K\"ahler form $J$ associated with the non-harmonic forms; however,
once we include these modes into the effective action, consistency of the supergravity action requires that we consider also the possibility of non-zero VEVs of these modes. Such VEVs  deform the Calabi-Yau $\hat Y_4$ into a non-K\"ahler manifold $\hat Z_4$. The supergravity reduction is thus really carried out on a non-K\"ahler space $\hat Z_4$, even though the F-theory vacuum corresponds to vanishing VEVs of the deformations and is thus defined in terms of the Calabi-Yau $\hat Y_4$ (or rather its singular blow-down $Y_4$).

With this understanding, it is important to specify precisely the geometries 
$\hat Z_4$ which induce the correct gauging in the F-theory effective action context.
The manifold $\hat Z_4$ has the same base $B_3$ 
as the resolved Calabi-Yau fourfold $\hat Y_4$, and all modifications 
take place in the resolution of the fibration. More precisely, $\hat Z_4$ is obtained as a non-K\"ahler resolution replacing \footnote{The explicit construction 
of $\hat Z_4$ should be the higher-dimensional generalization of the non-K\"ahler threefolds considered in \cite{Grimm:2010gk}.}
\beq
    \hat Y_4\ \rightarrow \ \hat Z_4\ .
\eeq
The mass terms for the scalars parameterizing this deviation from the Calabi-Yau constraint
depend on the geometric data of the resolved fourfold $\hat Z_4$. Due to the modifications in the 
elliptic fibration they involve the dilaton in accord with the Type IIB orientifold masses. 

Specifying our treatment of the geometrically massive $U(1)$s and the associated supergravity gauging requires specifying the non-harmonic forms referred to above: A set of non-closed two-forms $\{\tw_{0A}\}$ is required to describe the massive $U(1)_A$ potentials in the M-theory reduction of $C_3$ on $\hat Z_4$. Being non-closed these come with a set of three-forms $\{\alpha_a,\beta^a\}$ which intersect in a manner reminiscent of the symplectic structure on $H^3({X}_3)$ of a Calabi-Yau threefold. Consistency in turn implies the existence also of a set of dual four-forms $\{\tilde \tw_{aA}, \tilde \tw^{aA}\}$. The latter play a role in the F-theory  description of some of the gauge fluxes known to be present in a perturbative limit.
We now collect these forms and their intersections and then proceed to a  justification of  the intersection pattern.

We start with the structure required to describe the massive $U(1)$ bosons, i.e. a set of three-forms $\{\alpha_a,\beta^a\}$, two-forms $\{\tw_{0A}\}$, and four-forms $\tilde{\tw}^{bA}$ satisfying 
\beq 
\label{dtw}
d\tw_{0A} =  N_A \, C^a_A \, \alpha_a \;, \qquad \quad d \beta^a = - \delta^{ac} \frac12 \cK_{\alpha cb} \, N_A \, C^\alpha_A \, \tilde{\tw}^{bA} \;.
\eeq 
In these expressions we have introduced the constant matrices $C_A^a$ and $C^\alpha_A$ which for now are arbitrary but will be shown later to correspond precisely to the wrapping numbers (\ref{wrappingnum}) of the D7-brane in the IIB limit. The integers $N_A = \mathrm{rk} G_A$ are determined by the rank of the gauge group along the singularity, which we take to be $SU(N_A)$ in order to study the type IIB limit.  
Dimensional reduction on $\hat Z_4$ including the 2-forms 
$\tw_{0A}$ leads to additional three-dimensional vector and scalar modes
which are identified in the F-theory uplift with the gauge fields 
rendered massive in Type IIB by the St\"uckelberg mechanism. 
The index $0$ denotes that they are not in the Cartan resolution tree of the non-Abelian singularity associated to an index $A$.\footnote{The index $A$ need not be associated to a non-Abelian singularity but can also stand by itself corresponding to a single brane.} 

The intersection numbers  $\cK_{\alpha ab}$ appearing in (\ref{dtw}) are defined by evaluating the integrals 
$\int_{\hat Z_4} \omega_\alpha \wedge M \wedge N$, with $M,N \in \{\alpha_a,\beta^b\}$.
Here and in the sequel we write all integrals in terms of the non-K\"ahler space $\hat Z_4$; however, since the non-harmonic forms are present already on $\hat Y_4$, we could just as well evaluate these intersections on the Calabi-Yau $\hat Y_4$, finding identical expressions.
Due to the elliptic fibration structure of $\hat Z_4$ the $\left\{\alpha_a,\beta^a\right\}$ can be chosen to obey 
\beq
\label{intersectionNumbers}
 \int_{\hat Z_4} \omega_\alpha \wedge \alpha_a \wedge \alpha_b = 0 \ , \qquad
 \int_{\hat Z_4} \omega_\alpha \wedge \alpha_a \wedge \beta^b = \frac12 \cK_{\alpha ac} \delta^{cb} \ , \qquad
 \int_{\hat Z_4} \omega_\alpha \wedge \beta^a \wedge \beta^b = 0 \ ,  
\eeq
as well as
\beq
\label{tw_alpha_beta}
\int_{\hat Z_4} \eta \wedge \alpha_a \wedge \alpha_b = \int_{\hat Z_4} \eta \wedge \alpha_a \wedge \beta^b = \int_{\hat Z_4} \eta \wedge \beta^a \wedge \beta^b = 0 \ , \qquad \textnormal{with} \ \eta = \tw_{0A},\ \omega_0 \ .
\eeq
Here $\omega_\alpha$ are the basis elements in $H^2(\hat Z_4,\mathbb{R})$
which are obtained by pullback from the base $B_3$. Note that due to the factor $\tfrac12$ the $\cK_{\alpha ac}$ in \eqref{intersectionNumbers} are even and as we will see they correspond to the intersection numbers \eqref{kappaOnCY} defined on the double cover. The rationale behind this pattern is that $(\alpha_a, \beta^b)$ arise as the wedge product of 1-forms $(dx, dy)$ in the elliptic fiber and 2-forms in the base which pick up a minus sign under $SL(2,\mathbb Z)$ monodromies associated with the part of the discriminant locus that corresponds, in the Type IIB limit, to the orientifold plane. The set of forms $(\alpha_a, \beta^b)$ therefore inherits the symplectic structure of the 1-forms $(dx, dy)$ in the fiber.
Compatibility with \eqref{dtw} implies that one also has to demand 
\beq
\label{tw0-tildetw-omega}
C^\alpha_A \cK_{\alpha a b} \int \omega_\beta \wedge \tw_{0 B} \wedge \tilde{\tw}^{b A} = \delta^A_B \cK_{\beta a b} C^b_A \ ,\qquad C^\alpha_C \cK_{\alpha a b} \int \tw_{0A}\wedge \tw_{0 B} \wedge \tilde{\tw}^{b C} = 0 \ .
\eeq

We next turn to the four-forms required to include some of the gauge fluxes present in the perturbative Type IIB limit.
To implement these fluxes into F-theory we will need to introduce a set of four-forms $\tilde \tw_{aA}$. These are defined as the duals of the four-forms  $\tilde \tw^{a A}$ introduced in (\ref{dtw}) so that 
\bea \label{tildetw_tildetw}
&&    \int_{\hat Z_4} \tilde \tw_{aA} \wedge \tilde \tw^{bB} = \delta_A^B \delta_a^b, \\ \nonumber
&&    \int_{\hat Z_4} \tilde \tw_{aA} \wedge \tilde \tw_{bB} = - \frac12 \delta_{AB} \cK_{\alpha a b} N_A  C^\alpha_A. 
\eea 
The precise form of the intersection numbers in the second line of eq (\ref{tildetw_tildetw}) is dictated by the M2/D3-brane tadpole correspondence, as will be shown later.
The new four-forms satisfy the differential relations
\beq 
\label{dtwodd}
d\tilde \tw_{aA} =  N_A C^\alpha_A \omega_\alpha \wedge \alpha_a \; .
\eeq 
Again we can use partial integration and show the consistency of \eqref{tildetw_tildetw} with 
\eqref{dtw} and \eqref{intersectionNumbers}.

To complete the data required in a Kaluza-Klein reduction let us also give the remaining intersection numbers 
involving the non-closed forms. These can be deduced by requiring consistency with the Type IIB orientifold limit.
The forms $\tw_{0A}$ combine with the resolution forms $\tw_{iA}$
introduced in \eqref{twset} into the set 
\beq
\label{setI}
  \tw_{IA} \ ,\qquad I=0,i, \quad\quad i=1,\ldots,\text{rank}(G_A) \ , \quad\quad A = 1,\ldots, n_7\ ,
\eeq
where $n_7$ is the number of 7-brane stacks. 
Together they obey the intersection relations
\beq \label{twtw_Cartan}
  \int_{\hat Z_4} \tw_{IA} \wedge \tw_{JB} \wedge \omega_\alpha \wedge \omega_\beta
    = - \frac12 \delta_{AB}\, \cC^B_{IJ}\, C^\gamma_A \,  \cK_{\alpha \beta \gamma},   
\eeq
where  we are using the intersection numbers
\bea
\label{kappaOnBase}
\frac12 \cK_{\alpha \beta \gamma} = \int_{B_3}  \omega_\alpha \wedge \omega_\beta \wedge \omega_\gamma =  \int_{\hat Z_4} \omega_0 \wedge \omega_\alpha \wedge \omega_\beta \wedge \omega_\gamma.
\eea
Furthermore $\cC^A_{IJ}$ reduces to the Cartan matrix upon restriction of the indices to the Cartan algebra and $\cC^A_{0i}=0$, $\cC^A_{00} = N_A$. Note that \eqref{kappaOnBase} coincides with the definition of the intersection numbers given in \eqref{kappaOnCY} due to the fact that $\int_{X_3}  \omega_\alpha \wedge \omega_\beta \wedge \omega_\gamma = 2 \int_{B_3}  \omega_\alpha \wedge \omega_\beta \wedge \omega_\gamma$.

Finally,
\bea
\label{2tw0-tildetw}
\int_{\hat Z_4} \tilde \tw_{aA} \wedge \tw_{0B} \wedge \omega_\alpha &=& - \frac12 \delta_{AB} \cK_{\alpha a b} N_A C^b_A \; .
\eea
Let us reiterate that all intersection numbers defined here are integer-valued. The factors of $\frac12$ are chosen to allow the identification of $\cK_{\alpha a b}$ with the even-valued intersection numbers \eqref{kappaOnCY} on the double cover.

Having specified a basis for the dimensional reduction we can summarize the new 
fields completing the spectrum discussed in section \ref{sec:massless_reduction}.
Again, for simplicity, we will restrict to examples with 
$h^{(2,1)}(B_3)=0$ since the resulting bulk $U(1)$s do not play a role in the forthcoming discussion.
More precisely, we consider the expansion
\beq \label{C3_expand}
  C_3 = A^0 \wedge \omega_0 + A^\alpha \wedge \omega_\alpha + A^{I A} \wedge \tw_{IA} + c^a \alpha_a + b_a \beta^a  \,
\eeq
where $(A^0,A^\alpha , A^{I\, A})$ are vectors and $c^a, b_a$ are real scalars in three dimensions. 
As in the previous section the vectors $(A^0,A^\alpha , A^{I A})$ combine with 
real scalars $(v^0,v^{\alpha},  v^{IA})$ and fermions into supermultiplets.
Hence one needs to identify the geometric origin of the real scalars $v^{0A}$. Note that 
due to supersymmetry \textit{both} the vector and scalar fluctuations $A^{0A}$, $v^{0A}$ 
have to appear in the reduction. As we will argue momentarily the scalar fluctuation can be interpreted 
formally as a fluctuation into a non-K\"ahler space $\hat Z_4$.

As is well-known 
from the literature on compactifications on non-Calabi-Yau manifolds (see e.g.~\cite{Grana:2005jc} for a review), 
the condition on the existence of a supersymmetric effective theory is weaker than the condition to be in 
a supersymmetric vacuum. 
As in the threefold case the existence of a supersymmetric 
effective theory is expected if $\hat Z_4$ admits a globally defined $(1,1)$-form $J$.\footnote{We will 
not change the complex structure part of our geometry when going from $\hat Y_4$ to $\hat Z_4$.} 
Also the scalars $v^{0A}$ arise in the expansion of $J$ as
\beq \label{nonJ_expand}
   J =  v^0 \omega_0  + v^\alpha \omega_\alpha + v^{I A}  \tw_{I A}    \ .
\eeq
Note that in general $J$ is not a K\"ahler form since it is non-closed
\beq
dJ =  v^{0A} N_A C_A^a \alpha_a \;. 
\eeq 
It is well known that the supersymmetry conditions imply $dJ=0$ in the vacuum \cite{Becker:1996gj} and hence require 
\be
\left<v^{0A}\right> C_A^a = 0 \;. 
\ee
Therefore, while the vacuum configuration is a K\"ahler manifold, the geometric origin of the vector multiplet  
$(A^{0A},v^{0A})$ can be matched to (massive) resolutions of this manifold which take it away from K\"ahlerness.

Indeed from the perspective of the F-theory dual, as shown in section \ref{sec:MNonKaehler}, the $v^{IA}$ form a component of the 
four-dimensional vectors and so must have vanishing VEV by Lorentz invariance. The scaling in the F-theory limit is analogous to 
the one of the Cartan $U(1)$'s given in \eqref{F-limit_scalings}, i.e.
\beq \label{xi0-scaling}
   \xi^{0A} \equiv \frac{v^{0A}}{\cV} \ \propto \ \epsilon^{2}\ ,
\eeq
where the $\epsilon \rightarrow 0$ limit yields a hierarchy among the terms in the F-theory limit.
However, it will be crucial to keep the fluctuations $v^{0A}$ and $A^{0A}$  
in the spectrum of the effective theory. In the later sections we will argue that 
the F-theory lift of the three-dimensional vector multiplets $(v^{0A},A^{0A})$ yields the 
massive $U(1)$s encountered in the orientifold picture.\footnote{More precisely, they correspond to $2\pi\alpha' A^A_{(IIB)}$ in the orientifold limit.}

So far we have simply given the details of our proposal for implementing the massive $U(1)$s. The fact that the $U(1)$s arise from forms that are not harmonic (if the $C_A^a$ are non-zero) means that they are not massless. We present much more non-trivial checks of the proposal by studying the IIB limit in sections \ref{sec:d5tad} and \ref{sec:sugeff}.

It is worth noting that our proposal differs slightly in its implementation of the differential structure from similar setups such as presented in \cite{Gurrieri:2002wz,Grimm:2008ed,Camara:2011jg}. Once the first relation in (\ref{dtw}) is introduced the natural complement relation involves a set of non-closed 5-forms rather than the 3-forms we have introduced. More precisely we can define the set
\be
d\tw^{(2)}_{0A} = \alpha^{(3)}_A \;,\;\; d\beta^{(5)A} = \tilde{\tw}^{(6)0A} \;, \label{tw6}
\ee
such that 
\be
\int_{\hat{Z}_4} \tw^{(2)}_{0A} \wedge \tilde{\tw}^{(6)0B} = - \int_{\hat{Z}_4} \alpha^{(3)}_A \wedge \beta^{(5)B} = \delta_A^B \;.
\ee
Indeed such a structure is present in our setup as discussed in section \ref{sec:u1geometry}. Then, as usually happens, the first part of (\ref{tw6}) would correspond to reducing the electric field, in our case $C_3$, while the second part would encode the same physics but in the magnetic frame which in our case corresponds to reducing $C_6$. We only need to keep one of these descriptions which in our case is the electric one. Our implementation of the structure (\ref{tw6}) in (\ref{dtw}) amounts to expanding the non-closed 5-forms as a product of harmonic base 2-forms $\omega_{\alpha}$ and non-closed 3-forms $\beta^a$ which have a leg in the fiber. The non-closedness of the 3-forms is then inherited from that of the 5-forms. However the 3-forms $\beta^a$ are not the magnetic duals of the $\alpha_a$ in the sense of Hodge duality as the 5-forms were. Therefore they correspond to independent physics degrees of freedom and even in the electric frame we should still reduce $C_3$ on them as we do.\footnote{Of course the $\beta^a$ are magnetic duals to the $\alpha_a$ but in the sense of the $SL(2,{\mathbb Z})$ action.}

For later use we note that in accord with supersymmetry in the three-dimensional theory one expects that the $(c^a,b_a)$ combine into complex 
scalars as 
\beq \label{def-Na}
  N^a = c^a - i f^{ab}\, b_a \ .
\eeq
In the Calabi-Yau case, the complex structure for this combination is naturally induced
by the complex structure of $\hat X_4$. Formally generalizing this to the space $\hat Z_4$
one obtains the $N^a$ from reducing $C_3$ on complex 
$(2,1)$-forms $\Psi_a$ given by
\beq 
\label{Psi_exp}
\Psi_a =  \tfrac{i}{2} \R f_{ab} (\beta^b  - i   \bar f^{bc} \alpha_c) \ , \qquad \Psi_a + \bar \Psi_a = \alpha_a \;.
\eeq
The coefficient function $f^{ab}$ has 
to be chosen such that $\Psi_a$ are $(2,1)$-forms, and hence is a complex function $f^{ab}$ of the complex structure moduli of $\hat Z_4$. 
Here we denoted by $\R f_{ab}\equiv (\R f^{ab})^{-1}$ the inverse of the real part of $f^{ab}$.
Note that for harmonic forms $\Psi_a$ one can show that $f^{ab}$ can be chosen to be 
holomorphic in the complex structure deformations \cite{Grimm:2010ks}. We expect that it is 
possible to extend this structure to the forms $\Psi_a$ defined in \eqref{Psi_exp}.

\subsection{Abelian $G_4$ fluxes and their induced D5- and D3-tadpole}
\label{sec:d5tad}

In this section we begin to justify our framework for a  description of massive $U(1)$ symmetries in terms of the non-harmonic forms introduced in the previous section. Our first goal is to analyse the associated gauge fluxes and show their consistency with known results in the Type IIB limit. Apart from serving as an important check of our framework this leads us to a proposal for the F-theory uplift of chirality inducing gauge fluxes from Type IIB, which sheds more light on the nature of gauge fluxes in terms of the M-theory $G_4$ flux.

In M/F-theory gauge flux is encoded in suitable components of four-form flux $G_4$. Instead of presenting our proposal right away we  begin with some heuristics:
As reviewed in section \ref{gaugingIIB}, in the Type IIB limit the gauge fluxes descending from the ambient space can be expanded into flux quanta $\tilde{\cF}_I^{A, \alpha}$ along elements of $H^{1,1}_+(X_3)$ and $\tilde{\cF}_I^{A, a} $ along elements of $H^{1,1}_-(X_3)$. The two-forms in $H^{1,1}_+(X_3)$ uplift to two-forms on the base, i.e. to the elements $\omega_\alpha$. 
In fact, independently of any Type IIB considerations, it is well-known that  the Cartan fluxes in F-theory take the simple
form $G_4 =  - \sum_i \tilde{\cF}_i^{A, \alpha} \wedge \tw_{iA}$ with $\tw_{iA}$ given in eq. (\ref{twset}). 
This includes the uplift of orientifold even Cartan fluxes descending form the ambient space.
A natural guess is to extend this to the diagonal $U(1)$ flux by writing  $G_4 =  - \tilde{\cF}_0^{A, \alpha} \wedge \tw_{0A}$ for the non-harmonic two-form $\tw_{0A}$ introduced in (\ref{dtw}) (cf. also (\ref{setI})), and we will verify the validity of this ansatz momentarily.
On the other hand, not all fluxes must be given by four-forms expressible as the wedge product of two two-forms.
In particular this turns out to be the case for the uplift of the orientifold odd fluxes $\tilde{\cF}_I^{A, a} $ associated with the diagonal $U(1)$. Heuristically, this can be seen by noting that
the negative 2-forms  lift to 3-forms in F-theory, as will be recalled in more detail below. The non-harmonic three-forms $\{\alpha_a, \beta_a\}$ appearing, according to  eq. (\ref{dtw}), in the context of the massive $U(1)$ in turn are related to the set of non-harmonic four-forms $\{\tilde \tw_{aA}, \tilde \tw^{aA}\}$ of (\ref{tildetw_tildetw}). We will find that the $\tilde \tw_{aA}$ indeed have the right properties to describe such fluxes.

Altogether these considerations result in the following proposal for the form of the gauge fluxes associated with the massive $U(1)$s in F-theory,
\be
G_4 = - \tilde{\cF}_0^{A, \alpha} \omega_{\alpha} \wedge \tw_{0A}  - \tilde{\cF}_0^{A, a} \tilde \tw_{aA} . 
\label{newfluxes}
\ee

An important check of any ansatz for the $G_4$-form gauge flux is to compare its induced tadpoles with the well-known induced D5- and D3-tadpoles in the Type IIB limit.
Let us first consider uplifting the flux-induced D5-tadpole (\ref{d5tad1}) to F-theory. 
The uplift of the closed-string field-strengths locally takes the form of 
\be
F_3 \wedge dy \rightarrow G_4 \;,\;\; H_3 \wedge dx \rightarrow G_4 \;, \label{fluxup}
\ee
where the respective one-forms $dx$ and $dy$ correspond to the $A$- and $B$-cycle of the generic non-singular elliptic fiber. 
Both of these are odd under the orientifold monodromy, which means we can consider uplifting the 
odd two-forms $\omega_a$ to some 'even' three-forms by fibering them in a similar way
\be
\omega_a \wedge dy \rightarrow \alpha_a \;, \;\; \omega_a \wedge dx \rightarrow \beta^a \;. \label{omup}
\ee
The fact that such a fibration leads to globally well-defined forms can be deduced by considering 
the four-dimensional supergravities: the axions resulting from reducing $C_3$ on $\alpha_a$ 
and $\beta^a$ are the uplifts of the axions coming from reducing $C_2$ and $B_2$ 
on the $\omega_a$ (see section \ref{sec:sugeff} for much more detail).

We can use these relations to uplift the D5-tadpole constraint to a constraint on $G_4$-flux. Note however that it is dangerous to directly use \eqref{omup} to compare the coefficients in a form expansion, due to the difference in the intersection forms of $B_3$ and $X_3$. However, the integrated tadpole constraints are uplifted as
\be
\label{upliftD5tad}
\int_{X_3} d F_3 \wedge \omega_b \rightarrow \int_{\hat Y_4} d G_4 \wedge \beta_b.
\ee
In accord with our remarks before (\ref{intersectionNumbers}) we evaluate the integral right away on the Calabi-Yau resolution $\hat Y_4$ (ass opposed to the non-K\"ahler space $\hat Z_4$) describing the F/M-theory vacuum.

It is now evident that the gauge flux $G_4$ for the diagonal $U(1)$ flux must indeed be associated with non-harmonic forms, for which the above expression is non-zero.
In fact, using the non-closedness of $\tw_{0A}$ and $\tilde \tw_{aA}$ as given in \eqref{dtw} and (\ref{dtwodd}) one evaluates 
\beq
  dG_4 =  - N_A \Big(\tilde{\cF}_0^{A, a} C^\alpha_A + \tilde{\cF}_0^{A, \alpha} C^a_A \Big) \, \omega_\alpha \wedge \alpha_a\,. 
\eeq
This can be integrated to a global tadpole constraint 
\beq
\label{M5tad}
  \delta_{ab} \int_{\hat Y_4} dG_4 \wedge \beta^b =  - \frac12 N_A \Big( \tilde{\cF}_0^{A, c} C^\alpha_A + \tilde{\cF}_0^{A, \alpha} C^c_A  \Big) \cK_{\alpha ac} = 0 \;,
\eeq
in perfect agreement with the Type IIB result \eqref{D5tadInt}.
Hence such flux arises from expanding $G_4$ in two-forms $\omega_{0A}$ and four-forms  $\tilde \tw_{aA}$ which are not closed, 
precisely matching our claim. Indeed the fact that the wrapping numbers are the coefficients 
which control the non-closedness is exactly recreated in section \ref{sec:sugeff} from the gauged supergravity analysis.

Similarly we can match the D3-brane charge induced by diagonal $U(1)$ gauge flux in type IIB orientifold and in F-theory.
The expression (\ref{d3iibNgauge})  is to be compared with the flux contribution $\frac12 \int_{\hat Y_4} G_4 \wedge G_4$ appearing in the F/M-theory D3/M2-brane cancellation condition
\bea
N_{M2} + \frac12 \int_{\hat Y_4} G_4 \wedge G_4 = \frac{\chi(\hat Y_4)}{24}.
\eea

Evaluating this quantity for $G_4$-form flux of the type (\ref{newfluxes}) requires the intersection forms (\ref {twtw_Cartan}),   (\ref{tildetw_tildetw}), and (\ref{2tw0-tildetw}). The result is
\bea
\tfrac12 \int_{\hat Y_4} G_4 \wedge G_4 =  - \tfrac{1}{4} \sum_A N_A \Big( \cK_{\alpha \beta \gamma} C^\alpha_A \tilde \cF_0^{A, \beta} \tilde \cF_0^{A, \gamma} +   \cK_{\alpha b c} C^\alpha_A  \tilde \cF_0^{A,b}  \tilde \cF_0^{A, c} + 2    \cK_{a b \gamma} C^a_A \tilde \cF_0^{A, b}  \tilde \cF_0^{A, \gamma}             \Big)
\eea
and perfectly matches the Type IIB expression \eqref{d3iibNgauge}. In particular, this provides evidence for the intersection numbers (\ref{tildetw_tildetw}).

Before we proceed a comment on the quantisation of the non-harmonic fluxes is in order.
As is well-known, harmonic $G_4$-fluxes on a Calabi-Yau fourfold are subject to Witten's quantisation condition \cite{Witten:1996md}
\bea
\label{Witten-cond}
\tilde G_4 = G_4 - \frac14 p_1(\hat Y_4) \in H^4(\hat Y_4, \mathbb Z)\ , 
\eea
where $p_1(\hat Y_4)$ is the first Pontryagin class of $\hat Y_4$ which obeys $p_1(\hat Y_4) = - 2 c_2(\hat Y_4)$ for a Calabi-Yau fourfold.
This to be read as an equation in cohomology, or, equivalently, as the constraint that the integral of the left-hand side over an integral basis ${\cA_k} $ of four-cycles be integer-valued
\beq \label{rewrite_quantization}
   \int_{\cA_k}   \tilde G_4 = \int_{\hat Y_4} \tilde G_4 \wedge [\cA_k]  \quad \in \  \mathbb{Z} \ , \qquad \cA_k \in H_4(\hat Y_4,\mathbb{Z})\ ,
\eeq
where $[\cA_k]$ is the Poincar\'e dual class of $\cA_k$.
Eq. (\ref{Witten-cond}) is the analogue of the Freed-Witten quantisation  condition \cite{Freed:1999vc} for brane fluxes in Type II theories (see \cite{Collinucci:2010gz} for a recent analysis). 

Note that the condition \eqref{Witten-cond} cannot simply be generalized to the non-harmonic fluxes (\ref{newfluxes}), since the non-closed expansion forms $\omega_{\alpha} \wedge \tw_{0A} $ and $ \tilde \tw_{aA}$ are not elements inside the cohomology classes.
This cannot mean, though, that the non-harmonic fluxes are not subject to any quantisation condition. This is clear e.g.~by considering the uplift of $U(1)$ fluxes of concrete Type IIB models along the lines and extending the analysis of \cite{Collinucci:2008zs}. In F-theory models with a given Type IIB limit the quantization of the fluxes can of course be inferred from the perturbative Freed-Witten condition. A natural generalization of \eqref{rewrite_quantization} therefore is likely to involve 
a replacement of $[\cA_k]$ with a relative form in $H^4(\hat Y_4 ,D_A,\mathbb{Z})$ with integral coefficients. Such relative forms also can contain 4-forms 
which are exact on $\hat Y_4$ but have non-vanishing integral with a non-closed $\tilde G_4$. To make this more precise it will be 
crucial to specify the geometric part in $\tilde G_4$, the first Pontryagin class, which is likely replaced by $p_1(\hat Z_4)$ on the non-K\"ahler space. 
 A detailed specification of the quantisation condition directly in the language of the four-fold is however beyond the scope of this article and left for future studies.

\subsection{The induced chirality}
\label{sec:iibfluxchi}

One of the most important consequences of switching on gauge flux along the massive $U(1)$ symmetries is that the flux induces non-zero chirality
in the massless spectrum of charged matter states.
Such matter arises either as bulk matter propagating along the full divisor $D^{\rm b}_A$ in the base or as localised matter at the intersection curve of two 7-branes.
One of the strongest pieces of evidence why it is important to allow for gauge flux in the massive $U(1)$ is that the resulting chiral index in Type IIB orientifolds is protected as we take  the F-theory uplift. Indeed in Type IIB orientifolds, simple formulae for the chiral index of such massless matter in presence of gauge flux exist. In this section we make a proposal for the corresponding chirality formulae for the new type of F-theory fluxes, which precisely match the Type IIB results provided we  accept the intersection numbers introduced in section \ref{nonKaehlergeometry}.

At the intersection of two brane stacks $A$ and $B$ with (massive) Abelian symmetries $U(1)_A$ and $U(1)_B$ massless matter arises with relative Abelian charges
$(1_A, -1_B) + {\rm c.c}$ or  $(1_A, 1_B) + {\rm c.c}$. For later convenience we call matter curves of the first type $C_{AB}$ and matter curves of the second type $C_{AB'}$.\footnote{This is well familiar from experience with Type IIB orientifolds. If one uses the convention that the relative charge normalisation $(1_A, -1_B) + {\rm c.c}$ occurs at the  intersection of brane $A$ and $B$,  then the second type of matter is localised at intersections of brane $A$ with the orientifold image $B'$.} While the overall normalisation of the charges is of course conventional, the relative charges must be assigned such as to correctly reproduce the pattern of Yukawa couplings at codimension-3 singularity enhancements. 
This is a global feature of the geometry sensitive to the relative orientation of the branes at the two types of intersection loci.


It turns out that the chiral index for both situations can be consistently formulated in terms of the four-forms $\tilde{\tw}^{a A}$ introduced in section \ref{nonKaehlergeometry} and the two combinations 
\bea
\label{tildeDAnew}
[\tilde D_A] = [D_A] - \sum_{i=0}^{{\rm rk}(G_A)} a_{iA} \, \tw_{iA}, \quad\quad\quad 
[\tilde D_{A'}] = [D_A] - \sum_{i=0}^{{\rm rk}(G_A)} a_{iA'} \, \tw_{iA}.
\eea
Here $a_{iA} = a_{iA'}$, $i=1, \ldots, {\rm rk}(G)$, denote the Dynkin labels of the Dynkin diagram of gauge group $G_A$ and $a_{0A}= \frac{1}{N_A} = - a_{0A'}$ multiplies the non-harmonic 2-forms $\tw_{0A}$ introduced previously. For comparison with Type IIB we take $G_A=SU(N_A)$ with $a_{iA}=1$. The expression  $[D_A]$  denotes the 2-form dual to the pullback of the base divisor $D^{\rm b}_A$ as given in (\ref{DA}). The relative sign between $a_{0A}$ and $a_{0A'}$ sets the normalisation of the $U(1)_A$ charges.

In this subsection we concentrate on the contribution to the chirality of the fluxes along the diagonal $U(1)$, which as we have seen are associated to the non-harmonic forms $\tw_{0 A}$, $\tilde{\tw}_{a A}$. Our aim is to write a formula in terms of the geometric M-theory quantities that reproduces the correct type IIB chirality formula. The claim is then that the chirality $I_{AB}$ along matter locus $C_{AB}$ of a pair of branes $D_A$ and $D_B$ is given by the expression
\bea
\label{index1New}
I_{AB} = \frac{1}{4 }  \int_{\hat Z_4} \Big( ( [\tilde D_A] \wedge [\tilde D_{B'}]- [\tilde D_{A'}] \wedge [\tilde D_{B}]) + \cK_{\alpha a b} ( C^\alpha_A C^a_B \tilde{\tw}^{b A} - C^\alpha_B C^a_A  \tilde{\tw}^{b B}) \Big) \wedge G_4.
\eea
Note the appearance of the objects  $[\tilde D_{A'}]$, $[\tilde D_{B'}]$.
On the other hand, the quantity $I_{AB'}$ counting chiral matter along $C_{AB'}$ is given by
\bea
\label{index2new}
I_{AB'} = \frac{1}{4 }  \int_{\hat Z_4} \Big( ([\tilde D_A] \wedge [\tilde D_{B}]- [\tilde D_{A'}] \wedge [\tilde D_{B'}] ) - \cK_{\alpha a b} ( C^\alpha_A C^a_B  \tilde{\tw}^{b A} + C^\alpha_B C^a_A \tilde{\tw}^{b B} ) \Big) \wedge G_4.
\eea

We can evaluate the integrals appearing in these expressions using the formulae \eqref{twtw_Cartan}, \eqref{2tw0-tildetw} and \eqref{tw0-tildetw-omega}.
With these relations one finds after inserting $G_4$ given in \eqref{newfluxes} and neglecting Cartan fluxes
\bea
\label{index1eval}
I_{AB} &=& - \frac{1}{4 }  \Big( \cK_{\alpha \beta\gamma}C^\beta_A C^\gamma_B + \cK_{\alpha a b} C^a_A C^b_B \Big) \Big( \tilde{\cF}_0^{A,\alpha} - \tilde{\cF}_0^{B,\alpha} \Big) \\
&& - \frac{1}{4 }  \Big( \cK_{\alpha a b}C^\alpha_A C^a_B + \cK_{\alpha a b} C^a_A C^\alpha_B \Big) \Big( \tilde{\cF}_0^{A,b} - \tilde{\cF}_0^{B,b} \Big) \ .
\eea
Similarly
\bea
\label{index2eval}
I_{AB'} &=& - \frac{1}{4}  \Big( \cK_{\alpha \beta\gamma}C^\beta_A C^\gamma_B - \cK_{\alpha a b} C^a_A C^b_B \Big) \Big( \tilde{\cF}_0^{A,\alpha} + \tilde{\cF}_0^{B,\alpha} \Big) \\
&& - \frac{1}{4}  \Big( \cK_{\alpha a b}C^\alpha_B C^a_A - \cK_{\alpha a b} C^a_B C^\alpha_A \Big) \Big( \tilde{\cF}_0^{A,b} - \tilde{\cF}_0^{B,b} \Big) \ . 
\eea
This precisely matches the results one obtains from the corresponding type IIB expressions, where the chirality indices are given by\footnote{See appendix \ref{app:conventions} for more details regarding the IIB chirality. }
\bea
I_{AB} &=& - \int_{X_3} [D_A]\wedge [D_B] \wedge  (\tilde \cF_0^A - \tilde \cF_0^B) \ , \\
I_{AB'} &=& - \int_{X_3} [D_A]\wedge [D_B'] \wedge  (\tilde \cF_0^A + \sigma^* \tilde \cF_0^B) \ .
\eea

Even though we are focusing in this paper on fluxes associated with massive $U(1)$s the above chirality formula is expected to generalise, \emph{mutatis mutandis},
also to the case of massless $U(1)$s. A global prescription for their understanding was given in the $U(1)$ restricted Tate model of \cite{Grimm:2010ez} and it will be interesting to apply the above reasoning to constructions of this type.

\section{The M- and F-theory supergravity effective action}
\label{sec:sugeff}

In this section we derive the effective four-dimensional supergravity resulting from compactifications of F-theory on CY fourfolds including the non-K\"ahler deformations that arise from the non-closed forms introduced in section \ref{sec:abeff}. We do this by studying the reduction of M-theory to three-dimensions in subsection \ref{sec:MNonKaehler}, and then considering the F-theory limit in subsection \ref{F-theorylimit}. This will lead us to a rather complete picture of the effective 
$\cN=1$ supergravity theories which arise in an F-theory compactification including massive $U(1)$'s arising from fluxed and geometric gaugings. 
Furthermore we are able to to check that the resulting supergravity matches what we expect from the type IIB setting 
thereby providing further evidence towards the validity of our constructions.

\subsection{On the M-theory reduction on non-K\"ahler fourfolds}
\label{sec:MNonKaehler}

It is not hard to show that the vectors $(v^{0A},A^{0A})$
are indeed gauging scalar fields and can become massive by 
`eating' these scalars. 
To see that we compute the field strength of the reduction of $C_3$. 
With the help of \eqref{dtw} one finds 
\beq
  F_4 = F^\Lambda \wedge \omega_\Lambda + \nabla  c^a \wedge \alpha_a 
  + db_a \wedge \beta^a + b_a \, d\beta^a\ ,
\eeq
where we introduced the abbreviation $\omega_\Lambda = (\omega_0,\omega_\alpha,\tw_{IA})$, and similarly 
the field-strengths $F^\Lambda$. In particular we see that the scalars  $c^a$ appear with a covariant derivative given by 
\beq \label{Dca}
  \nabla c^a = d c^a -  N_A C^a_A A^{0A} \ .
\eeq
This structure is of course very reminiscent of what we found in \eqref{gauging1} in 
the context of the geometric St\"uckelberg mechanism for D7-branes in Type IIB language.
The would-be shift symmetry of the scalars $c^a$ is gauged by the gauge potentials $A^{0 A}$.

Clearly, the covariant derivative \eqref{Dca} forces the complex scalars $N^a$ defined in \eqref{def-Na}
to be gauged via the covariant derivative  
\beq \label{nablaN}
   \nabla N^a = d N^a -  N_A C^a_A A^{0A}\ .
\eeq

In this section we develop the 3D $\cN=2$ supergravity that we should match the M-theory reduction to. In \cite{Berg:2002es} the action is given for the case where either all the fields are scalars and some are gauged or where all the gauged fields are dualised to vectors. For the M-theory reduction we require an action where some of the gauged scalars are dualised to vectors while some gauged scalars are left as scalars. We perform the explicit dualization in appendix \ref{dualization_appendix} and only display the 
result here.
Let us consider chiral multiplets $M^I$\footnote{The index $I$ labeling these chiral fields is not related to the index enumerating the two-forms $\tw_{I A}$. We trust that the appropriate range for the index I will always be clear from the context.} and vector multiplets $(\xi^{ \Lambda},A^{ \Lambda})$.
We denote the field strength of $A^{ \Lambda}$ by $F^{\Lambda}$. Any three-dimensional gauged ${\cal N}=2$ supergravity with this field content can then be cast into the form
\bea\label{kinetic_lin_gen_1}
  \cS^{(3)}_{\cN=2} &=& \int \Big[ -\tfrac{1}{2}R_3 *1 - 
  \tilde K_{I \bar J }\, \nabla M^I \wedge * \nabla \bar M^{J}
  + \tfrac{1}{4} \tilde K_{\Lambda \Sigma}\, 
  d\xi^{\Lambda}\wedge * d\xi^{\Sigma} \\ 
  && \phantom{\int \Big[} + \tfrac{1}{4} \tilde K_{\Lambda \Sigma}\, F^{\Lambda} \wedge * F^{\Sigma}
     + \,  F^{\Lambda} \wedge \I (\tilde K_{\Lambda I} \, \nabla M^I) 
     + \tfrac12 \Theta_{\Lambda \Sigma} A^{\Lambda} \wedge F^{\Sigma} - (V_\cT + V_{\rm F}) * 1 \Big]\; , \nn
\eea
with covariant derivatives
\beq \label{DMTheta}
   \nabla M^I = d M^I + X^I_\Lambda \, A^{\Lambda} \;.
\eeq
The scalar potential in the action \eqref{kinetic_lin_gen_1} is 
given by 
\bea \label{3dPotential}
   V_\cT &=& \tilde K^{I \bar J} \cT_I \cT_{\bar J} - \tilde K^{\Lambda \Sigma} \cT_\Lambda \cT_\Sigma - \cT^2, \\ 
   V_{\rm F} &=&  e^K (\tilde K^{I \bar J} D_I W \overline{D_J W} - (4 + \xi^\Sigma \xi^\Lambda \tilde K_{\Sigma \Lambda} ) |W|^2) \; , \nn 
\eea 
where $\cT_I = \partial_{M^I} \cT$, $\cT_\Lambda =  \partial_{\xi^\Lambda} \cT$, and $\tilde K^{\Lambda \Sigma}$ and $\tilde K^{I \bar J}$ are the inverses of $\tilde K_{\Lambda \Sigma}, \tilde K_{I \bar J}$, 
respectively.

The three-dimensional action is thus specified by a 
kinetic potential $\tilde K(M,\bar M| \xi)$, which depends on the 
complex scalars $M^I$ and the real scalars $\xi^\Lambda$ in the 
vector multiplets. $\tilde K$ determines 
the kinetic terms as 
\beq
    \tilde K_{I \bar J } = \partial_{M^I} \partial_{\bar M^J} \tilde K\ , \qquad   
    \tilde K_{I \Lambda} = \partial_{M^I} \partial_{\xi^\Lambda} \tilde K \ ,\qquad 
    \tilde K_{\Lambda \Sigma} = \partial_{\xi^\Lambda} \partial_{\xi^\Sigma} \tilde K \ , 
\eeq
and the K\"ahler covariant derivative $D_I W = \partial_{M^I} W + (\partial_{M^I} \tilde K) W$.
Furthermore one has to specify the constant `embedding tensor' $\Theta_{\Lambda \Sigma}$,
the Killing vectors $X^{J}_\Lambda$ in \eqref{DMTheta},   
as well as the function $\cT(M,\bar M,\xi)$.
To determine the function $\cT$ one first has to evaluate 
\beq \label{3dDterm}
   i \partial_{M^I} D_\Sigma = \tilde K_{I \bar J} X^{\bar J}_\Sigma\ ,
\eeq
just as in the four-dimensional theory \eqref{genD}.
The resulting potentials $D_\Sigma$ appear in $\cT$ as 
\beq \label{explicitcT}
   \cT = -  \tfrac12 \xi^\Sigma \Theta_{\Lambda \Sigma}  \xi^\Lambda +  \xi^\Sigma D_\Sigma\ .
\eeq
Using \eqref{3dDterm} one readily evaluates for constant Killing vectors
\bea \label{generalVcT}
  V_\cT &=& -(\tilde K_{I \bar J}-\tilde K_{\Gamma I}\tilde K^{\Gamma \Delta}\tilde K_{\Delta \bar{J}}) X^I_\Sigma  X^{\bar J}_\Lambda \xi^\Sigma \xi^\Lambda - \tilde K^{\Lambda \Sigma} D_\Lambda D_\Sigma \\
   && -  \Theta_{\Sigma \Gamma} \tilde K^{\Gamma \Delta} \Theta_{\Delta \Lambda} \xi^\Sigma \xi^\Lambda + 2 \tilde K^{\Lambda \Sigma} D_\Lambda \Theta_{\Sigma \Gamma} \xi^\Gamma  + 2i\tilde K^{\Lambda\Sigma}\tilde K_{\Lambda I} X^{I}_{\Delta} \Theta_{\Sigma \Gamma} \xi^\Delta \xi^\Gamma \nn \\&&
  - 2i\tilde K^{\Lambda\Sigma}\tilde K_{\Lambda I} X^{I}_{\Delta} D_{\Sigma}\xi^\Delta - \cT^2 
  \ . \nn 
\eea

Note that starting with the action \eqref{kinetic_lin_gen_1} one can also dualize the 
vector multiplets $(A^\Sigma,\xi^\Sigma)$ into complex scalars $t_\Sigma$. In order 
to determine the K\"ahler potential $K$ for all scalars $t_\Sigma,M^I$ one 
applies a Legendre transform
\beq \label{LegendreK}
   \text{Re}\, t_\Sigma = \partial_{\xi^\Sigma} \tilde K\ , \qquad K(t,\bar t,M,\bar M) = \tilde K - \tfrac12 (t_\Sigma + \bar t_\Sigma) \xi^{\Sigma}\ , 
\eeq 
as discussed in appendix \ref{dualization_appendix}. Due to the non-trivial 
$\Theta_{\Sigma \Lambda}$ the dual scalars $t_\Sigma$ are gauged as 
\beq
  \nabla t_\Sigma = dt_\Sigma - 2 i \Theta_{\Sigma \Lambda} A^\Lambda\ .
\eeq
This completely specifies the relevant three-dimensional supergravity theory. We now turn to matching this to the M-theory reduction.

Let us now specify the fields and couplings in the general action \eqref{kinetic_lin_gen_1}
when we reduce M-theory on the non-K\"ahler space specified in subsection \ref{nonKaehlergeometry}.
One first notes that the vector multiplets are identified as 
$(\xi^\Lambda , A^\Lambda) = \big\{ (R,A^0),\ (L^\alpha, A^\alpha),\ (\xi^{IA},A^{IA}) \big\}$,
where $(A^0,A^\alpha,A^{iA})$ are the vectors appearing in \eqref{C3_expand} and we 
have to set 
\beq \label{RLxi_def}
   R = \frac{v^0}{\cV} \ , \qquad L^\alpha = \frac{v^\alpha}{\cV} \ , \qquad \xi^{IA} = \frac{v^{IA}}{\cV} \ ,
\eeq
with $\cV$ being the volume of $\hat Z_4$, and $(v^0,v^\alpha,v^{IA})$ are 
the coefficients in \eqref{nonJ_expand}. Furthermore, we identify the 
complex scalar fields $M^I = \big\{ N^a , z^\cK \}$,
where $N^a$ was given in \eqref{def-Na} and $z^\cK$ are the complex structure deformations 
of $\hat X_4$.

The kinetic potential $\tilde K(M,\bar M|\xi)$ determining the dynamics of the fields $(\xi^\Lambda , A^\Lambda) $ and $M^I$ 
has to be computed by inserting the Kaluza-Klein Ansatz of section \ref{nonKaehlergeometry} into the 
the eleven-dimensional supergravity action. This will lead to explicit expressions for the 
kinetic term, which can be used to deduce the $\tilde K$. Clearly, this is more involved than in 
the Calabi-Yau case due to the appearance of the non-closed forms. Compared with the Calabi-Yau reductions of 
\cite{Haack:1999} one expects the replacement of the appropriate intersections as introduced in section \ref{nonKaehlergeometry}.
Formally this yields the expression 
\beq
\label{kinPot}
\tilde{K}(M,\bar M|\xi) = -3\log \cV + \frac{i}{4}\xi^{\Lambda} (N^a - \bar{N}^a)(N^b- \bar{N}^{b})\int_{\hat Z_4}\omega_{\Lambda}\wedge \Psi_a \wedge\bar{\Psi}_{\bar{b}} + {K}_{CS}\ , 
\eeq
where the volume $\cV$ has to be expressed as a function of the $(\xi^\Lambda , A^\Lambda) = (R,L^\alpha,\xi^{IA})$ defined in \eqref{RLxi_def}, and $\Psi_a$ are the complex structure 
depended $(2,1)$-forms introduced in \eqref{Psi_exp}.
The part of the kinetic potential $\tilde{K}_{CS}=-\log \int \Omega \wedge \bar \Omega$ corresponding to the complex structure moduli can be obtained from a dimensional reduction of the curvature scalar, but this will not be needed in this paper. Note that the kinetic potential depends only on the imaginary parts of the $N^a$ consistent with the fact that the real parts enjoy a gauged shift symmetry according to eq. \eqref{nablaN}. Note that in order 
to explicitly check \eqref{kinPot} one needs to generalize the techniques developed for non-Calabi-Yau threefolds (see \cite{House:2004pm} for example). In particular, one has to show how the Hodge-star can be evaluated 
on the non-harmonic forms introduced in section \ref{nonKaehlergeometry}. However, crucial for our analysis will be that the non-harmonic forms induce a scalar potential. 

Having determined the kinetic terms in the dimensionally reduced effective action 
we turn now to the discussion of the scalar potential.  
The vector multiplet gauging can be read off from the Chern-Simons term of 11-dimensional supergravity
\beq
   S_{CS}^{(11)} = -\frac{1}{12} \int C_3 \wedge F_4 \wedge F_4 =  \int_{M_{2,1}} \Theta_{(IA) \alpha} A^{IA} \wedge F^\alpha +\frac{1}{2}\Theta_{\alpha \beta} A^\alpha \wedge F^\beta +\ldots \ ,
\eeq 
where we will focus on the terms with flux dependent constant coefficients $\Theta_{(IA) \alpha} , \Theta_{\alpha \beta}$ etc. 
Evaluating the internal integral, one finds
\be \label{Theta1}
\Theta_{(IA)\alpha} = - \frac12 \int_{\hat Z_4} \tw_{IA} \wedge \omega_\alpha \wedge G_4 \ ,
\ee
where $G_4$ includes the new fluxes introduced in \eqref{newfluxes} in addition to standard harmonic fluxes. Note that only $\Theta_{(0A)\alpha}$ receives a contribution from the second term due to the non-closedness of the $\tw_{0A}$.
Besides, fluxes which lift to F-theory have to obey
\beq \label{Theta2}
  \Theta_{\alpha \beta} = - \frac12 \int_{\hat Z_4} \omega_\alpha \wedge \omega_\beta \wedge G_4 = 0\ , \qquad \Theta_{(IA)(JB)} = - \frac12 \int_{\hat Z_4} \tw_{IA} \wedge \tw_{JB} \wedge G_4 = 0 \ ,
\eeq
and similarly for the other index combinations. It is our working assumption that such intersection properties can be achieved in the F-theory limit and possibly even in the full M-theory  by a suitable redefinition of the basis of forms or a shift of the fluxes.\footnote{This is indeed possible in a harmonic reduction as discussed in
\cite{GrimmSavelli}.}
The non-trivial fluxes thus determine the first term of $\cT$ given in \eqref{explicitcT} as 
\beq \label{cTflux}
  \cT^{\rm flux} = -  \xi^{IA} \Theta_{(IA)\alpha} L^{\alpha} =  \frac{1}{ 4 \cV^{2}} \int_{\hat Z_4} J \wedge J \wedge G_4\ ,
\eeq
where for the last equality we have used \eqref{Theta2} and \eqref{nonJ_expand}. This formally agrees with the flux potential found in a harmonic reduction. However, 
due the non-closed forms in $G_4$ one finds new contributions. Let us denote the part of $\cT^{\rm flux}$ induced by the new
forms in $G_4$ by $\cT^{\rm flux}_{U(1)}$. The first derivative of this potential is used in 
the scalar potential and can be evaluated by inserting \eqref{newfluxes} into \eqref{cTflux} as
\bea \label{Dterm_flux}
   D^{\rm flux}_{0A} \equiv \partial_{\xi^{0A}} \cT^{\rm flux}_{U(1)} &=&  - L^\alpha \tilde{\cF}_0^{B, \beta}  \int_{\hat Z_4} \tw_{0A} \wedge \omega_\alpha \wedge  \omega_\beta \wedge \tw_{0B} \\
 && -  L^\alpha \tilde{\cF}_0^{B, a}  \int_{\hat Z_4} \tw_{0A} \wedge \omega_\alpha \wedge  \tilde \tw_{a B} \nn \\ 
 &=& \frac12 \cK_{\alpha \beta \gamma} N_A C_A^\gamma L^\alpha \tilde{\cF}_0^{A, \beta}    +  \frac12 \cK_{\alpha ab} N_A C^b_A L^\alpha \tilde{\cF}_0^{A, a}    \ , \nn
\eea
where we have used \eqref{twtw_Cartan} and \eqref{2tw0-tildetw}.
In the next subsection we will lift this term to four space-time dimensions and show 
that it corresponds to a D-term arising from a gauged $U(1)$ symmetry.

The gaugings of the chiral multiplets are obtained by comparing the 
covariant derivatives \eqref{nablaN} with the general expression \eqref{DMTheta} such that 
\be \label{killingVectorX}
X^{a}_{0A} = - N_A C^{a}_{A} = X^{\bar{a}}_{0A}\ , \qquad X^{a}_{\alpha} = 0 = X^{a}_{iA}, \ i\neq 0 \;.
\ee
Note that these Killing vectors are constant so that one can integrate \eqref{3dDterm}
to obtain $D_{0A} =  i C^a_{A}\tilde K_{\bar{N}^a}$ for the only nonzero potential. 
From the gauged supergravity expression \eqref{3dPotential} 
we can read off the resulting scalar potential. 
Using the general expression \eqref{explicitcT} together with \eqref{Theta1}, \eqref{Theta2} and \eqref{kinPot} one finds
\bea  \label{complete_cT}
\cT  &=&  - \xi^{iA} \Theta_{(iA)\alpha} L^{\alpha} +  i \xi^{0A} N_A C^a_{A}\tilde K_{\bar{N}^a}  \\
   & = & \frac{1}{ 4\cV^{2}} \int_{\hat Z_4} J \wedge J \wedge G_4  + \frac{1}{ 4\cV^{2}} \int_{\hat Z_4} J \wedge J \wedge d C_3 \ ,\nn 
\label{evaluatedcT}
\eea
where we have used the expression \eqref{kinPot} for the kinetic potential to determine the first derivative $\tilde K_{N^a}$.
Let us stress that the scalar potential computed using $\cT$ directly contains a mass term for the massive diagonal $U(1)$ 
in the F-theory limit.  This $U(1)$ comes from the vector multiplet $\left(A^{0A},\xi^{0A}\right)$ which also contains 
the scalar $\xi^{0A}$. As we will see in the next subsection the second term in \eqref{complete_cT} induces a mass term 
for $\xi^{0A}$ when inserted into the scalar potential \eqref{generalVcT}.

\subsection{The F-theory limit}
\label{F-theorylimit}

In the previous section we have studied the three-dimensional effective 
theory arising in the reduction of M-theory on a non-K\"ahler Calabi-Yau 
fourfold. In a next step we aim to lift these results to four dimensions 
and show that they reduce to the orientifold effective action discussed 
in section \ref{gaugingIIB} in the weak coupling limit.

To study the F-theory limit we dimensionally reduce a general four-dimensional  
$\cN=1$ supergravity action involving a set of vectors $A^{IA}$ and chiral scalars $M^n=\{t_{\alpha},N^a,z^\cK\}$ to three dimensions and compare it to the 
action obtained by compactifying M-theory on the non-K\"ahler 
fourfold $\hat Z_4$. 
We will focus on the couplings which capture the 
gaugings of the shift symmetries. In particular, the gaugings induce 
covariant derivatives for the gauged scalars and a D-term potential
in the general action \eqref{eq:N_1} as
\beq \label{S4gauge}
  \cS^{(4)}_{\rm gauge} = \int - K_{m \bar n} \nabla M^m \wedge *_4  \nabla \bar M^{\bar n} - \tfrac12 \R f^{(IA) (JB)} D_{(IA)} D_{(JB)} *_4 1\ , 
\eeq
where the covariant derivatives $\nabla M^n = d M^n + X_{IA}^n A^{IA}$ split as
\beq \label{non-trival_covariant}
    \nabla t_{\alpha} = d t_{\alpha} + X_{\alpha IA} A^{IA} \ ,\qquad  \nabla N^a = d N^a + X_{IA}^a A^{IA} \ ,
    \qquad \nabla z^\cK = dz^\cK \ .
\eeq
The Killing vectors $X_{IA}^I$, $X_{\alpha IA}$ are constant for a gauged shift symmetry.

In the dimensional reduction to $d=3$ one splits the four-dimensional metric $g^{(4)}_{\mu \nu}$ 
and its inverse as 
\beq \label{metric_red}
  g^{(4)}_{\mu \nu} = \left(\begin{array}{cc} g^{(3)}_{rs} + R^{-1} A^0_r A^0_s & R^{-1} A^0_r \\ 
                                               R^{-1} A^0_r & R^{-1} \end{array} \right)\ ,\qquad 
  g_{(4)}^{\mu \nu} = \left(\begin{array}{cc} g_{(3)}^{rs} & - g_{(3)}^{rs} A^0_s  \\ 
                                                - g_{(3)}^{rs} A^0_s  & R + g_{(3)}^{rs} A^0_r A^0_s \end{array} \right)\ ,  
  \eeq 
where $r$ is the circumference of the 4th dimension as in \eqref{def-R}.  
Note that this also allows us to derive the simple split of the determinant $\det g^{(4)} =  r^2 \cdot  \det g^{(3)} $.
Furthermore, one splits the four-dimensional vectors as
\beq \label{AAred}
  A^{IA} = (A^{IA}_3  - R^{-1} \xi^{IA} A^0_3,\ R^{-1} \xi^{IA})\ , 
\eeq
and $A^0_3$ is the vector in the reduction of the four-dimensional metric. The vector 
$A^0_3$ combines with $R$ into a three-dimensional vector multiplet.
The $\cN=2,d=3$ vector multiplets then contain the bosonic 
fields $(\xi^{IA},A_3^{IA})$, and $(R,A_3^0)$.
Inserting the decomposition \eqref{AAred} into \eqref{S4gauge}, and imposing 
that the $M^n=\{t_{\alpha},N^a,z^\cK\}$ are independent of the 4th dimension one finds 
\bea \label{S3gauge}
  \tilde{\cS}^{(3)}_{\rm gauge} &=& \int - K_{m \bar n} \nabla M^m \wedge *_3  \nabla \bar M^{\bar n} - V_{\rm gauge} *_3 1\ ,\\
  V_{\rm gauge} &=& K_{m \bar n} X_{IA}^m X_{JB}^{\bar n}\xi^{IA} \xi^{JB}  + \frac12 R \cdot \R f^{(IA) (JB)} D_{IA} D_{JB}\ ,\nn
\eea
with three-dimensional covariant derivatives  
\beq \label{3dgauge-der}
  \nabla M^m = d M^m + X_{IA}^m A^{IA}_3 \ .
\eeq
In this expression we have rescaled the three-dimensional metric as $g_3 \rightarrow R^{-1} g_3$ to 
bring the Einstein-Hilbert term into its standard $d=3$ form.
It is important to remark that the vector $A^0$ cancels in the gauging \eqref{3dgauge-der} due to the appearance 
of $A^0$ in the metric ansatz \eqref{metric_red}.

The action can be rewritten in the standard three-dimensional form of \eqref{kinetic_lin_gen_1}. The kinetic terms of the vectors $(\xi^{IA},A_3^{IA})$ and the scalars $N^m$ are encoded by the kinetic potential
\beq \label{Kinetic_after_red}
   \tilde \cK(M,\bar M|\xi,R) = \log R + K(M,\bar M) - \frac{1}{R} \R f_{(IA)(JB)} \xi^{IA} \xi^{JB}\ , 
\eeq
which is obtained from the four-dimensional K\"ahler potential $K(N,\bar N)$ and gauge-coupling 
function $f_{(IA)(JB)}(N)$. This implies that the inverse of the kinetic metric is given by 
\beq \label{tKinverse}
   \tilde \cK^{00} = -R^2 \ , \qquad \tilde \cK^{(IA)(JB)} = -  \frac12 R\, \R f^{(IA)(JB)} - \xi^{IA} \xi^{JB} \ ,\qquad
   \tilde \cK^{(IA)0} = -  R\xi^{IA}\ ,
\eeq
where the index $0$ labels the $R$-direction. As the embedding tensor satisfies $\Theta_{(IA)(JB)}=0$, equation \eqref{explicitcT} reduces to $\cT = -D_{IA} \xi^{IA}$. Using \eqref{tKinverse} one can then explicitly check that the potential $V_{\rm gauge}$  in \eqref{S3gauge} takes the standard form
\be
V_{\rm gauge} = \tilde \cK^{m \bar n} \cT_m \cT_{\bar n} - \tilde \cK^{(IA) (JB)} \cT_{IA} \cT_{JB} - \cT^2.
\ee
Hence one concludes that the additional term proportional to $\xi^{IA} \xi^{JB}$ in 
\eqref{tKinverse} precisely 
ensures that the $-\cT^2$ term cancels for a kinetic potential $\tilde K$ of the form determined by \eqref{Kinetic_after_red} and \eqref{legendreTrf} in accord with the positive definiteness of the four-dimensional potential.

We would now like to uplift the three-dimensional action obtained in section \ref{sec:MNonKaehler} by comparing it to the general expression \eqref{S3gauge} one obtains by reduction from four dimensions. 
The four-dimensional K\"ahler potential and gauge kinetic function can be obtained from the three-dimensional kinetic potential using \eqref{Kinetic_after_red}. However, $\tilde \cK$ is not identical to the kinetic potential obtained in section \ref{sec:MNonKaehler}, where the complex scalars $t_{\alpha}$ had been dualized into vector multiplets $(L^{\alpha}, A^{\alpha})$. As detailed in appendix \ref{dualization_appendix}, the potentials are related by a Legendre transformation
\beq
\label{legendreTrf}
 \tilde \cK(M,\bar M|\xi^{IA},R) = \tilde{K}(z,N| \xi^{\Lambda}) - \tfrac12 (t_{\alpha}+\bar{t}_{\alpha})L^{\alpha}, \qquad\qquad \R \  t_{\alpha} = \tilde{K}_{L^\alpha},
\eeq
where $\tilde K$ is a function of the complex scalars $M^n=\{t_{\alpha},N^a,z^\cK\}$ as well as the real vector multiplet 
scalars $\xi^{IA},R$.
Unfortunately for the kinetic potential obtained in eq.~\eqref{kinPot} the relation $\R \ t_{\alpha} = \tilde{K}_{L^{\alpha}}$ cannot be explicitly inverted, so that we only obtain an implicit expression for the four-dimensional K\"ahler potential $K$ from \eqref{Kinetic_after_red} and \eqref{legendreTrf}. Explicitly the match of the 4d gauge coupling function and K\"ahler potential is performed as follows. One expands the large volume
expression for  $\cV$ in \eqref{kinPot}  keeping track of the $\epsilon$ scalings \eqref{F-limit_scalings}, \eqref{xi0-scaling}, and 
the form of the intersection numbers \eqref{twtw_Cartan}:\footnote{A more detailed discussion of this limit can be found in \cite{Grimm:2010ks,GrimmSavelli}.}
\bea \label{tKexpansion}
    \tilde{K}(M,\bar M|\xi) &=& \log R +  \log \Big(\frac12\frac{1}{3!} \cK_{\alpha \beta \gamma} L^\alpha L^\beta L^\gamma 
    - \frac{1}{4 R} \cC^A_{IJ} C_A^\lambda \cK_{\lambda \alpha \beta} L^\alpha L^\beta \xi^{IA} \xi^{JA}  + \cO(\epsilon^3) \Big) \nn \\
         &&+ \frac{i}{4} L^{\alpha} (N^a - \bar{N}^a)(N^b- \bar{N}^{b})\int_{\hat Z_4}\omega_{\alpha}\wedge \Psi_a \wedge\bar{\Psi}_{\bar{b}} + {K}_{CS} \ , 
\eea
where we have used the vanishing of the intersections $\int \tw_{IA} \wedge \Psi_a \wedge \bar \Psi_b = \int \omega_0 \wedge  \Psi_a \wedge \bar \Psi_b =0$ due to \eqref{tw_alpha_beta}.
The unusual factor of $\tfrac12$ in our expansion of the volume
\be
\label{volume_exp}
\cV = \int_{Z_4} \frac{1}{4!} J^4 = \frac12\frac{1}{3!} \cK_{\alpha \beta \gamma} v^\alpha v^\beta v^\gamma v^0 + ...
\ee
is a result of the intersection numbers \eqref{kappaOnBase}, which were chosen in this way in order to be able to identify them with the intersection numbers \eqref{kappaOnCY} on the double cover to facilitate the match with the IIB reduction.

Next one has to expand the logarithm in \eqref{tKexpansion} to quadratic order in $\xi^{IA}$, perform the 
Legendre transform \eqref{legendreTrf}, and compare the result with the expression \eqref{Kinetic_after_red} obtained from the 4d to 3d reduction. Note that in order that the 4d gauge coupling completes to 
\beq
\label{gaugeKin_M}
    f_{(IA) (JB)} = \frac{1}{2} \delta_{AB} \, \cC^A_{IJ} \, C_A^\alpha\ t_\alpha \ ,
\eeq 
one has to add a term proportional to 
$\frac{1}{R} \cC^A_{IJ} C_A^\lambda \xi^{IA} \xi^{JA}  (N^a - \bar{N}^a)(N^b- \bar{N}^{b})\int_{\hat Z_4}\omega_{\lambda}\wedge \Psi_a \wedge\bar{\Psi}_{\bar{b}} $ to the kinetic potential \eqref{tKexpansion} as discussed in \cite{Grimm:2010ks}.

To perform the match with the IIB reduction of section \ref{gaugingIIB} we identify the correspondence between the fields to be
\be
\label{mappingM-IIB}
v^\alpha \leftrightarrow v^\alpha_B \ , \qquad N^a \leftrightarrow G^a \ , \qquad A^{0A} \leftrightarrow 2\pi\alpha' A^A,
\ee 
where $v^\alpha_B$ denote the K\"ahler moduli of the base.
The gauging of the $N^a$ given in \eqref{nablaN} is not changed by the uplift from three to four dimensions and matches the gauging \eqref{gauging1} of the $G^a$ if one takes into account the relation of the gauge fields. To identify the relationship between $t_\alpha$ and the $T_\alpha$ defined in \eqref{def_Talpha} we use \eqref{tKexpansion} to find
\be
\R \  t_\alpha = \partial_{L^\alpha} \tilde{K}  = \frac{1}{4}\cK_{\alpha\beta\gamma} L^\beta L^\gamma R \cV^3 + ...
\ee
Now one uses the fact that in the F-Theory limit $\epsilon \rightarrow 0$ of vanishing fiber volume the K\"ahler moduli scale as \cite{Grimm:2010ks}\footnote{The factor of 2 again arises from the factor in the definition of the intersection numbers.}
\be
\label{scaling_L}
L^\alpha \rightarrow 2 \frac{v^\alpha_B}{\cV_B}.
\ee
At leading order in $\epsilon$ this implies
\be
R = \frac{2}{\cV^3} \Big(\frac{1}{3}\cK_{\alpha\beta\gamma}L^\alpha L^\beta L^\gamma\Big)^{-1} + ... \rightarrow \frac14 \frac{\cV_B^2}{\cV^3}.
\ee
We are therefore led to identify
\be
\label{relation_tT}
t_\alpha \rightarrow \frac{1}{4}\cK_{\alpha\beta\gamma} v_B^\beta v_B^\gamma + ... = \frac12 T_\alpha.
\ee
Note that under the identifications of the various fields as given above the gauge kinetic function \eqref{gaugeKin_M} agrees with the Type IIB expression \eqref{gaugeKinIIB}.

As shown in Appendix \ref{dualization_appendix} the gauging of the scalars $t_\alpha$ after the Legendre transformation is given in terms of the embedding tensor \eqref{Theta1} by
\be
\nabla t_\alpha = d t_\alpha -2i\Theta_{\alpha IA} A^{IA}.
\ee
Inserting the expansion of the $G_4$-flux and using the integrals \eqref{twtw_Cartan}, \eqref{2tw0-tildetw} one finds for the gauging with respect to the diagonal $U(1)$
\be
\Theta_{\alpha  0A} = -\frac{1}{4} N_A \Big( \cK_{\alpha\beta\gamma} \tilde{\cF}^{A,\beta} C^\gamma_A + \cK_{\alpha a b} \tilde{\cF}^{A, a} C^b_A \Big).
\ee
Using \eqref{relation_tT} this precisely reproduces the gauging \eqref{gauging2} found in the IIB reduction. 

In appendix \ref{dualization_appendix} it is shown that the Legendre transformation implies $\partial_{t_{\alpha}} K = \partial_{t_{\alpha}} \tilde{\cK} = -\frac12 L^\alpha$, while the derivatives with respect to the $N^a$ are not changed when going from the kinetic potential to the four-dimensional K\"ahler potential. Using this we can evaluate the four-dimensional D-terms despite not being able to carry out the Legendre transformation explicitly. From \eqref{kinPot} one obtains
\bea
D_{0A} &=& -i K_{\bar N^a}\bar{X}^a_{IA} + 2 K_{\bar T_\alpha} \Theta_{\alpha 0A} = i\tilde{K}_{\bar{N}^a} N_A C^a_A - \Theta_{\alpha 0A} L^\alpha \nn \\
&=& \frac{v^\alpha}{4\cV} N_A \Big( \cK_{\alpha\beta\gamma} \tilde{\cF}^{A,\beta} C^\gamma_A + \cK_{\alpha a b} (\tilde{\cF}^{A, a}-\delta^{ac}b_c) C^b_A \Big).
\eea
Taking account of the relations \eqref{mappingM-IIB} and \eqref{scaling_L} this matches the D-term of the corresponding type IIB theory given in \eqref{DtermIIB}.

Finally let us note that the scalar potential \eqref{generalVcT} encodes a mass term for the scalar $\xi^{0A}$ which becomes a component of the $U(1)$ gauge field upon uplifting to four dimensions. After setting fluxes to zero we find that the purely geometric contribution to this mass is given by the first term in \eqref{generalVcT}. Using the identities presented in appendix \ref{dualization_appendix} it is straightforward to check that this  reproduces the mass $m^2_{AB} \propto K_{a\bar b} C^a_A C^b_B$ obtained in the IIB reduction in \eqref{massIIB} after rescaling to obtain canonical kinetic terms.

The matching of the gaugings and the associated masses and D-terms derived in this section provide a further nontrivial check on our proposal regarding the description of massive gauge symmetries using non-harmonic forms as well as the F-theory uplift of IIB fluxes presented in section \ref{sec:d5tad}.

\section{The geometry of the $U(1)$s}
\label{sec:u1geometry}

In the previous sections we have argued purely within supergravity that expansion of $C_3$ in non-closed 2-forms correctly reproduces the effective action of massive $U(1)$ bosons as expected from a Type IIB perspective.
In this section we discuss in more detail the origin of these forms from a geometric point of view. 

To this end we start from the well-known geometric realisation of massless $U(1)$ symmetries and their non-Abelian generalisations in F/M-theory.
As described in section \ref{sec:abeff} the Abelian gauge bosons residing in the Cartan of a non-Abelian gauge group $G$ follow by Kaluza-Klein expansion of the M-theory 3-form $C_3$ along a set of ${\rm rk}(G)$ harmonic 2-forms $\tw_{i \, A}$ \footnote{We suppress the brane index $A$ in the sequel. These two-forms are the Poincar\'e dual of a set of holomorphic divisors $D_{i}$.}. 
On the other hand, the gauge bosons related to the non-Cartan generators are due to M2-branes wrapped along certain curves which
in the singular limit have vanishing volume such that the wrapped M2-branes become massless. 
There are two ways to make both the divisors $D_{i}$ associated with the Cartan $U(1)$s and the curves associated with the roots visible: by resolving or by deforming the singularity.

Let us start with the deformation of the singularity by moving the 7-branes supporting the non-Abelian gauge group $G$ off each other in the base. 
On a general Calabi-Yau 4-fold, the deformed 7-branes generically intersect along the curve of self-intersection of the brane divisor in the base after such a deformation. The fiber over this curve exhibits singularity enhancement. This complication does not occur for F-theory on K3, where the 7-branes are points on the base $B$. We therefore focus on this case to demonstrate the geometric origin of the membrane curves in the deformed phase.\footnote{A recent in depth analysis of the 2-cycle and group theory structure on elliptic K3 is provided in \cite{Braun:2008ua}.}
An important fact which we will use is that on K3 local mirror symmetry implies that the resolution and deformation are dual to each other \cite{Katz:1997eq}.

\subsection{Local deformations and resolutions}

Consider an $A_{N-1}$ singularity whose defining equation in ${\mathbb C}^3$ is 
\bea
y^2 = -x^2 + z^{N} \;.
\eea
The singularity is at $z=0$. The singularity admits a universal deformation which is the so-called preferred versal form 
\be
y^2 = -x^2 + z^{N} + \sum_{I=2}^{N} b_I z^{N-I} \;.
\ee 
We will generally denote indices that run up to $N$ with $I$ and indices that run up to $N-1$ with $i$. We can write the deformed singularity as 
\be
y^2 = -x^2 + \prod_{I=1}^{N} \left(z+t_I\right) \;,
\ee 
where the $b_I$ are then elementary symmetric polynomials of degree $I$ in the coordinates $t_I$ and we have to impose that
\be
b_1 = \sum_{I=1}^N t_I = 0\;. \label{tracelessness}
\ee
The $N$ parameters $t_I$ parameterise the deformation of the singularity such that if they are all non-vanishing the original singularity at $z=0$ is fully deformed. The singularity enhances at the points $z=-t_I$ up to the full $A_{N-1}$ for all the $t_I$ coincident. Therefore there are $N$ enhancement points. At each such point an $S^1$ collapses which we can identify explicitly by projecting to the imaginary plane of $(x,y)$ 
\be
\left(\I\; y\right)^2 + \left(\I\; x\right)^2 = -\prod_{I=1}^N \left(z+t_I\right) \;.
\ee
Within the K3, the collapsing $S^1$ is the A-type cycle of the elliptic fiber. One can construct 2-cycles stretched between the branes by fibering this collapsing $S^1$ over a real curve connecting the $t_I$ points. The curves $v_i$, with $i=1,...,N-1$, are the fibration of the collapsing $A$-cycle over the line $t_i - t_{i+1}$,
\bea
v_i: A{\rm -cycle} \, \,  \rightarrow \, (t_i - t_{i+1}).
\eea
The geometry matches the group theory of $SU(N)$ as follows: the $N$ $t_I$ are identified with the weights of the fundamental representation, while the $N-1$ $v_i$ are identified the simple roots. Each simple root is associated to a non-Cartan generator and this generator corresponds to an M2-brane wrapping the curve $v_i$. More general generators result from M2-branes along all possible chains of $v_i$, $C_{kl} = v_k \cup v_{k+1} \cup \ldots \cup v_l$, with both orientations of the M2-branes taken into account \cite{Katz:1996ht}.

Let us now turn to the resolution, which replaces the singularity in the fiber by a set of ${\rm rk}(SU(N)) =N-1$ homologically independent resolution $\mathbb P^1$s.
If the singularity is embedded into a compact space, these $N-1$ resolution $\mathbb P^1$ combine with the compliment of the resolution tree to form 
$N$ curves, only $N-1$ of which are homologically independent. 
On K3 the complexified K\"ahler parameters (measuring the volume and B-field) of these $\mathbb P^1$ are mirror dual to the deformation parameters $t_I$ and will be denoted by $\tilde t_I$.

The sum $\sum_{I=1}^N \tilde{t}_I$ is homologically trivial matching the relation of the weights (\ref{tracelessness}). The associated $N-1$ homology representatives are given by the dual to the $v_i$ which we denote $\tilde{v}_i= \tilde t_i - \tilde t_{i+1}$. These 2-cycles intersect according to the Cartan matrix as indeed the simple roots do.\footnote{
Together with $v_0 =  - \sum_i v_i$ they therefore form the nodes of the extended Dynkin diagram. 
If the singularity is embedded into the elliptic fiber of an elliptic fibration, as in the applications to F-theory, the 
group theoretic identity $\sum_{j=0}^{N-1} v_j =0$ translates into the homological relation $\sum_{j=0}^{N-1} [v_j] + [e] = [0]$
with $[e]$ denoting the class of the smooth elliptic fiber.} 

The Cartan $U(1)$s arise from expanding the M-theory 3-form $C_3$ in 2-forms $\tw_i$ which are associated to the root 2-cycles $\tilde{v}_i$ in the following way: To see the explicit relation we need to be careful regarding the intersection of the roots, and for this it is useful to consider the Calabi-Yau 4-fold case. Although in 4 dimensions two 2-cycles naturally intersect at a point, in 8 dimensions this applies to 6-cycles and 2-cycles. Then to preserve the Cartan matrix intersection structure we should define the inner product between two roots, or 2-cycles, by associating a dual 6-cycle to each root and then intersecting it with another root's 2-cycle.\footnote{See e.g.  \cite{Intriligator:1997pq}, where the curves and divisors are respectively denoted by  $\epsilon$ and $S$ in (8.18).} A way to associate these divisors, which we denote by $\tilde t^I$, is by taking a basis that satisfies
\be
\tilde{t}_I \cdot  \tilde{t}^J = -\delta_{I}^{J} \;.
\ee
Then the 2-forms Poincar\'e dual to the 6-cycles $D_i= \tilde{t}^{i}-\tilde{t}^{i+1}$ are the $\tw_i$.\footnote{To conform with our previous notation, we label the divisors $D_i$ with a downstairs index even though they correspond to the Cartan generators.} In the case of K3 the $D^I$ and $\tilde{v}^i$ happen to be again 2-cycles. Likewise the Poincar\'e duals of the 2-cycles $\tilde{t}_I$ are also 2-forms.

\subsection{The diagonal $U(1)$}

We now turn to the picture for the diagonal $U(1)$. Our proposal is that the diagonal $U(1)$ is accounted for by the fact that there are $N$ resolution spheres but only $N-1$ cohomology classes. To be more explicit consider the 2-forms $\omega_I$ Poincar\'e dual to the divisors $\tilde{t}^I$ so that 
\be
\int_{\tilde{t}_I} \omega_J = -\delta_J^I \;.
\ee
Then we have
\be
-1 = \int_{\sum_{I=1}^N \tilde{t}_I} \omega_J = \int_{\partial {\cal C}} \omega_J = \int_{\cal C} d\omega_J \;,
\ee
where we have introduced the chain ${\cal C}$ associated to the homological triviality of the $N$ spheres. This means that the $\omega_J$ cannot be closed. However $\tw_i = \omega_i - \omega_{i+1}$ is closed, harmonic, and leads to the Cartan $U(1)$s as discussed in the previous section.

The claim is that the diagonal $U(1)$ arises from expanding $C_3$ in the two form
\be
\label{tw0-prop}
\tw_0 \equiv \sum_{I=1}^N \omega_I \;.
\ee
Such a form is not closed and is naturally associated to the sum of the weights $\sum_{I=1}^N \tilde{t}_I$. 

This proposal holds irrespective of the dimension of the Calabi-Yau including the physically relevant case of 4-folds. However, certainly on K3, there is something missing from the picture because we know that for K3 the $U(1)$ should in fact be massless and so correspond to a harmonic form. This is clear from the type IIB perspective, where the brane and its image are in the same homology class. Therefore in the K3 case we expect a massless $U(1)$ and correspondingly a resolution parameter such that $\sum_{i=1}^N t_I \neq 0$. The key point to realise is that, as expected from IIB, the masslessness of this $U(1)$ is not a local property. To see how the $U(1)$ arises explicitly in a non-local sense it is simpler to consider the deformation picture. Consider starting from an $A_{N+M-1}$ singularity
\be
y^2 = -x^2 + z^{N+M} \;,
\ee 
and deforming it to an $A_{N-1}\times A_{M-1}$ singularity
\be
y^2 = -x^2 + \left(z+t^{(N)}\right)^{N}\left(z+t^{(M)}\right)^{M} \;,
\ee 
which can be further deformed completely as
\be
y^2 = -x^2 + \prod_{I=1}^{N} \left(z+t^{(N)}_I\right) \prod_{I=1}^{M} \left(z+t^{(M)}_I\right) \;.
\ee 
The initial deformation amounts to separating a stack of branes so that we break $SU(N+M)\rightarrow SU(N)\times SU(M) \times U(1)$. If we look locally at each singularity we would have the constraint 
\be
\sum_{I=1}^N t^{(N)}_I = \sum_{I=1}^M t^{(M)}_I = 0 \;.
\ee
However we know that in the initial singularity we only had the constraint 
\be
\sum_{I=1}^N t^{(N)}_I + \sum_{I=1}^M t^{(M)}_I = 0 \;.
\ee
Hence we are free to take a deformation with say $\sum_{I=1}^N t^{(N)}_I \neq 0$, but we should consider it in a global sense.  Similarly there is a new 2-cycle which we can wrap membranes on which cannot be seen locally but rather corresponds to the fibration of the A-cycle over the line $t^{(N)}-t^{(M)}$. 

The dual setup to this, involving resolutions such that $t_I \rightarrow \tilde{t}_I$, then amounts to precisely the case where the diagonal $U(1)$ is massless. And like the deformation this is a non-local property (on K3 the non-local part is the same since the mirror symmetry does not act on the trivial $R^8$ base).\footnote{The fact that a singularity resolution can be locally non-K\"ahler but not globally so due to homological relations is familiar from CY conifold transitions \cite{Greene:1995hu}.} Let us translate this deformation picture explicitly to the resolution discussion presented above. We would now have
\be
\sum_{I=1}^N \tilde{t}^{(N)}_I + \sum_{I=1}^M \tilde{t}^{(M)}_I = \partial {\cal C} \;,
\ee
however 
\be
0 = \int_{\partial {\cal C}} \left(M\tw^{(N)}_0 - N\tw^{(M)}_0 \right) = \int_{\cal C} d \left( M\tw^{(N)}_0 - N\tw^{(M)}_0 \right) \;,
\ee
which identifies the closed 2-form combination and massless mode.

The picture presented here has been for the $U(1)$ associated to a separation of brane stacks. However in the IIB picture we were concerned with a brane stack and its orientifold image. The same non-local geometry discussion follows through in that case as well. We present the explicit details in appendix \ref{Dnbreaking} by deforming a $D_N$ singularity to an $A_{N-1}$ one.

\section{Interpretation and Further Directions}
\label{sec:concl}

In this article we have made a proposal to describe the uplift of the diagonal $U(1) \subset U(N)$ gauge symmetries of Type IIB orientifolds to F-theory. In orientifold models with exchange involution, the diagonal $U(1)$ of a given brane stack is massive.
We have collected evidence based on  M/F-theory duality that such massive $U(1)$ bosons uplift to modes from expanding the M-theory 3-form $C_3$ in certain non-harmonic 2-forms. We have shown in detail, by compactifying M-theory to three dimensions, that this reduction ansatz correctly encodes the gauging of the axions participating in the St\"uckelberg mechanism and the St\"uckelberg mass itself. Including also non-harmonic 4-forms we have been able to extend this match to a proposal for the uplift of the D3-brane and D5-brane tadpoles, as well as the chirality formula from Type IIB to F-theory.

Our 3-dimensional supergravity analysis fits into the framework of supergravity reduction on non-K\"ahler manifolds. A reduction of the K\"ahler form $J$ along the non-harmonic 2-forms leads to a 3-dimensional scalar $v^{0A}$ 
whose VEV measures deviation from the K\"ahler condition $dJ=0$. However, in the vacuum $v^{0A} =0$, we recover the Calabi-Yau condition. Note that in the 3-dimensional M-theory compactifications the volumes $v^{0A}$ and the vectors $A^{0A}$ are in the same supermultiplet, and lift together to the components of the four-dimensional $U(1)$ vector comprising the diagonal $U(1)$. Clearly, in a 4-dimensional vacuum the diagonal $U(1)$ should have a zero VEV to preserve Poincar\'e invariance. However, for the fluctuations supersymmetry requires the use of non-harmonic forms both in 
the reduction of the supergravity forms and the forms characterizing the geometry. This is is familiar from other $\cN=1$
compactifications on non-Calabi-Yau threefolds. In these cases one obtains massive R-R gauge bosons from the 
reduction into non-harmonic forms, as discussed e.g.~in \cite{Grimm:2008ed}. These R-R gauge bosons become massive
by `eating' an R-R axion which sits with the non-Calabi-Yau deformations in the same supermultipet. 

The successful match of the defining data of the F/M theory and Type IIB supergravity therefore leads us to conjecture the presence of a special set of non-harmonic 2-forms $\tw_{0A}$ in the geometry of elliptically fibered Calabi-Yau fourfolds that describe the uplift of Type IIB orientifolds with exchange involution. Locally, using group theoretic arguments, we have argued that a natural candidate for this 2-form is the combination $(\ref{tw0-prop})$ of 2-forms dual to the $N$ resolution 2-cycles $\tilde t_I$ of an $A_{N-1}$ singularity. However, from the IIB setting we know that the mass of the $U(1)$ is not a local property but rather a global one. Although we were able to deduce many global properties of the non-harmonic forms by using the supergravity analysis we did not give a complete global geometric identification of such forms. Such an identification should involve a criterion for distinguishing the special set $(\ref{tw0-prop})$ from the infinity of other non-harmonic forms in the manifold. 
We will sketch two possible interpretations in the following.

A first possibility is to more rigorously establish a global treatment of the local arguments presented in section \ref{sec:u1geometry}. More precisely, one 
might be able to argue that the number of $U(1)$'s are captured by the precise number of holomorphic 
curves supported in the globally resolved elliptic fibration in the Calabi-Yau prescription. 
In fact, a basis of such holomorphic curves 
can be found for elliptic fibrations realized in toric ambient spaces. A direct computation shows that among the 
basis element for the Mori cone of holomorphic curves there is one generator corresponding to the 
extra curve \eqref{tw0-prop}. An independent variation of the volume of this curve will account 
for the extra fluctuation violating the Calabi-Yau condition. However, it remains to establish 
precise global criteria when it is possible to construct the associated higher non-harmonic forms 
which we used in our reduction. 
Most interesting would be to find quasi-topological 
data of the resolved fourfold which allow one to count the number and compute the couplings of such massive diagonal
$U(1)$s. One could  hope to find an analog of the Mordell-Weil group which can be 
studied for massless $U(1)$s. 

A second possibility would be that the relevant two-forms are given by elements in the torsion cohomology groups of the fourfold.
Recently, in \cite{Camara:2011jg}, such torsion forms were invoked in the context of the dimensional reduction of the closed string R-R forms of Type II orientifolds in order to describe massive R-R $U(1)$ forms. It was also suggested that massive type IIA D6 open-string $U(1)$s can be described in this way by uplifting to M-theory on a $G_2$ manifold. If this latter proposal is correct, it is natural to suggest that our non-closed forms can be counted by torsion cohomology of the resolved non-K\"ahler space $\hat{Z}_4$ or the resolved Calabi-Yau space $\hat{Y}_4$. It would be very interesting to make this conjecture more concrete with an explicit example (perhaps by extending the work of \cite{Collinucci:2008zs}). 

If the non-closed forms proposed in this work (and prior to that in \cite{Grimm:2010ez}) are indeed associated to torsion cohomology then there are interesting consequences. In this case there is an obvious associated discrete $\mathbb Z_k$ symmetry which manifests itself in the effective theory. 
In particular, the St\"uckelberg mechanism induces a breaking of the $U(1)$ gauge theory to a $\mathbb Z_k$ gauge theory, with $k$ given by the charge of the St\"uckelberg axion participating in the Higgsing \cite{Maldacena:2001ss,Gonzalez:2011wy}. This implies that we would expect massive solitonic objects with charge mod $k$ with respect to this $\mathbb Z_k$ gauge theory \cite{Maldacena:2001ss,Camara:2011jg}\footnote{In our context, a candidate for such objects would be M5-branes wrapped along torsional 5-cycles. These 5-cycles should have a component in the base which stretches from the brane to its image in the double cover picture.}, and also a remnant exact discrete symmetry in the effective action which may have useful phenomenological implications \cite{Gonzalez:2011wy}.
It is not clear what discrete symmetry should be associated to our constructions. We expect that for a single brane stack it is either trivial (which would be associated to a chain with one boundary) or $\mathbb Z_2$, and that the rank should increase with the number of brane stacks.  

One interesting aspect of the $U(1)$ symmetries studied is their effect on instantons.
In the orientifold limit it is well-known that the $U(1)$s, even when massive, remain as perturbative selection rules organising the pattern of operators in the effective theory (such as Yukawa couplings), and are broken only non-perturbatively by D3-brane instantons \cite{Blumenhagen:2006xt}. This includes the possibility of D3-D(1)-D(-1) bound states or equivalently of fluxed D3-instantons \cite{Grimm:2011dj}, whereas instantons with vanishing D3-charge do not contribute to the superpotential.  D3-brane instantons lift to M5-instantons in M-theory \cite{Witten:1996bn}, whose effects in F-theory have recently been under closer investigation in \cite{Blumenhagen:2010ja}. Note that M5-instantons are not accounted for by the geometry of the elliptic fibration. This gives rise to the expectation  \cite{Grimm:2011dj} that the $U(1)$ selection rules persist in M- and F-theory, violated only by corresponding M5-instantons, even though there will in general be no simple $g_s$ suppression any more for the mass of the gauge potential away from the strict orientifold regime. The fact that the $U(1)$ selection rules continue to operate in the F-theory uplift is corroborated by our supergravity analysis, which identifies  a method to trace back the dynamics of the $U(1)$ bosons.

Our analysis in this paper was based on F-theory models which have a simple type IIB uplift. On the other hand, it is also well-known that generic classes of the F-theory models exhibit a structure of Yukawa couplings beyond what is possible in Type IIB orientifolds due to exceptional singularity enhancement \cite{Donagi:2008ca}. In those models we expect that there is no diagonal $U(1)$, not even a massive one, associated with an $A_{N-1}$ singularity along a brane divisor. It would be interesting to investigate if this is the case and if so what is the geometric mechanism responsible for the disappearance of the non-harmonic 2-forms in this case. One expects that this is due to a more complicated monodromy structure because of exceptional enhancement. However, since the study of global monodromies 
can be notoriously complicated it would be nice to find a more direct topological criterion. 

\subsection*{Acknowledgments}

We gratefully acknowledge discussions with Federico Bonetti, Pablo Camara, Frederik Denef, Hirotaka Hayashi, Arthur Hebecker, Albrecht Klemm, Sven Krause, Peter Mayr, Christoph Mayrhofer, Hans Peter Nilles, Raffaele Savelli, Stefan Theisen and Wati Taylor.
TW would like to thank the MPI Munich, 
 EP is grateful to the MPI Munich and the ITP Heidelberg for hospitality during the preparation of this work.
The work of
EP was supported by the European ERC Advanced Grant 226371 MassTeV and
the PITN contract PITN-GA-2009-237920. The work of TW and MK was supported in part by SFB-Transregio 33 ``The Dark Universe'' by the DFG.

\appendix

\section{Conventions}
\label{app:conventions}
This appendix summarizes the conventions we use in the Type IIB orientifold setting. All Ramond-Ramond fields $C_p$ as well as the NS-NS 2-form $B_2$ are chosen dimensionless in the sense that $\int_{\Gamma_p} C_p$ is dimensionless for any p-chain $\Gamma_p$. The kinetic terms of the Ramond-Ramond fields in the democratic formulation are given by
\be
S_{kin} = - \frac{1}{8\kappa_{10}^2} \int G_p \wedge \ast G_p \ ,
\ee
where the field strengths are defined as $G_{p+1} = d C_p - d B_2 \wedge C_{p-2}$, $p= 0,2,4,6,8$. They satisfy the duality relations
\be
\ast G_p = (-1)^{(p-1)/2} G_{10-p},
\ee
which are to be imposed at the level of the equations of motion. Finally the gravitational coupling constants are dimensionless in these conventions
\be
\frac{1}{2\kappa_{10}^2} = \frac{1}{2\kappa_{4}^2} = 2\pi.
\ee

The Chern-Simons action of a single Dq-brane along a cycle $\Gamma$ of the Calabi-Yau manifold $X_3$ is given by
\be
S_{CS}= - 2 \pi  \, \int_{{\mathbb  R}^{1,3} \times \Gamma} \sum_{p} C_{2p} \wedge   {\rm tr}\left[ e^{2 \pi \alpha' \mathbf F }\right]      \,  \sqrt{\frac{\hat A(TD)}{\hat A(ND)}}\, .
\ee
Taking the orientifold projection into account we add the actions of brane and image-brane stacks and divide by two. In this picture an $O7$-plane carries a relative charge of -4 compared to the D7-branes. For the purpose of deriving D7-, D5- and D3-tadpoles we must include an additional factor of $\frac12$ in the Chern-Simons actions (but not in the Dirac-Born-Infeld action) when using the democratic formulation. This compensates for the fact that it includes explicitly both the equivalent electric and magnetic degrees of freedom.

To describe a stack of $N_A$ D7-branes along a divisor $D_A$ and the corresponding image stack along $D_A'$ it is convenient to define the combinations
\be
D_A^\pm = D_A \cup \pm D_A'.
\ee
The expansion of the respective Poincar\'e duals in terms of the coh
\be
[D_A^+] =  C^\alpha_A \omega_\alpha \ , \qquad [D_A^-] =  C^a_A \omega_a \ .
\ee

The index measuring the net chirality between two stacks of branes is given by
\bea
I_{AB} = -    \int D_A \wedge D_B \wedge  (\tilde{\cF}_0^A - \tilde{\cF}_0^B)
\eea
with $\tilde{\cF}_0^A $ defined in (\ref{tildeF}).
For the chirality along the intersection of a brane stack and an image stack one must further take into account that fluxes along an image stack are given by $\tilde{\cF}' = - \sigma^* \tilde{\cF}$, leading to
\bea
I_{AB'} = -   \int D_A \wedge D_B' \wedge  (\tilde{\cF}_0^A + \sigma^* \tilde{\cF}_0^B) \ , \\
I_{AA'} = -    \int D_A \wedge D_A' \wedge (\tilde{\cF}_0^A + \sigma^* \tilde{\cF}_0^A) \ .
\eea
Here $\tilde{\cF}$ includes a contribution from the Kalb-Ramond B-field as defined in \eqref{tildeF}.
In order to study the F-Theory lift of these quantities it is helpful to expand these expressions by inserting the expansions \eqref{frakfdef} of the fluxes into positive and negative parity basis forms. Using the intersection numbers \eqref{kappaOnCY} one straightforwardly finds
\bea
I_{AB} =& - \frac{1}{4 }  \Big( \cK_{\alpha \beta\gamma}C^\beta_A C^\gamma_B + \cK_{\alpha a b} C^a_A C^b_B \Big) \Big( \tilde{\cF}_0^{A,\alpha} - \tilde{\cF}_0^{B,\alpha} \Big) \nn \\
& - \frac{1}{4 } \Big( \cK_{\alpha a b}C^\alpha_A C^a_B + \cK_{\alpha a b} C^a_A C^\alpha_B \Big) \Big( \tilde{\cF}_0^{A,b} - \tilde{\cF}_0^{B,b} \Big) \ , \\
I_{AB'} =& - \frac{1}{4 }  \Big( \cK_{\alpha \beta\gamma}C^\beta_A C^\gamma_B - \cK_{\alpha a b} C^a_A C^b_B \Big) \Big( \tilde{\cF}_0^{A,\alpha} + \tilde{\cF}_0^{B,\alpha} \Big) \nn \\
& - \frac{1}{4 }  \Big( \cK_{\alpha a b}C^\alpha_B C^a_A - \cK_{\alpha a b} C^a_B C^\alpha_A \Big) \Big( \tilde{\cF}_0^{A,b} - \tilde{\cF}_0^{B,b} \Big) \ , 
\eea
where we have focused on the contribution due to fluxes along the diagonal $U(1)$. Note that the dependence on the continuous modulus $b^a$ in the expansion $B = B_+ + B_- = b^\alpha \omega_\alpha + b^a \omega_a$ has dropped out in accordance with the fact that the chirality index is a discrete quantity. Furthermore the discrete quantity $b^\alpha$ contributes only to $I_{AB'}$ and not $I_{AB}$. Finally, for the intersection of a brane stack with its own image stack the second formula simplifies to
\be
I_{AA'} = - \frac{1}{2} \Big( \cK_{\alpha \beta\gamma}C^\beta_A C^\gamma_A - \cK_{\alpha a b} C^a_A C^b_A \Big) \tilde{\cF}_0^{A,\alpha} \ .
\ee

\section{Dualising the 3D action} 
\label{dualization_appendix}

In this appendix we give more details regarding the dualisation of some of the gauged scalars to vectors in 
the general $\cN=2$ three-dimensional action. As discussed in section \ref{sec:abeff} this is required to match 
the three-dimensional action resulting from direct reduction of the M-theory action. We start from the general 
action for the gauged scalar multiplets given in \cite{Berg:2002es}
\beq\label{kinetic_lin_gen_1app}
  \cS^{(3)}_{\cN=2} = \int-\tfrac{1}{2} R_3 *1 - 
  K_{A \bar B }\, \nabla N^A \wedge * \nabla \bar N^{\bar B}  
     - \tfrac12 \Theta_{AB} A^{A} \wedge F^{B} - (V_\cT + V_F) * 1 \; , 
\eeq
with covariant derivatives
\beq \label{DMThetaapp}
   \nabla N^A = d N^A + \Theta_{BC}\tilde{X}^{A B} \, A^{C} \;.
\eeq
The gauging is implemented via this covariant derivative and yields the equations of motion for the 
vector fields $A^{A}$ given by 
\be
\label{eomA}
 * F^{A} = 2 \R \big( K_{B\bar{C}} \tilde{X}^{\bar{C} A} \nabla N^{B} \big) \;.
\ee
The scalar potential is given by
\beq
   V_\cT = K^{A \bar B} \partial_A \cT \partial_{\bar B} \cT - \cT^2\ ,\qquad V_F= e^K (K^{A \bar B} D_A W \overline{D_B W} - 4 |W|^2) \;.
\eeq 
Note that in the formulation \eqref{kinetic_lin_gen_1app} the scalars are the propagating degrees of freedom 
in the theory, while the vectors carry no propagating degrees of freedom. The object $\Theta_{AB}$ plays the role of the `embedding tensor' specifying which vectors appear in the gauging. Note that the following manipulations go through almost unchanged if in addition to the charged scalars one includes a set of propagating vector multiplets $(\tilde \xi^r, \tilde A^r)$. In this case the only alteration would be that index $M^I$ in the following formulae should run over $\{\tilde \xi^r, M^I\}$ instead.\\
The full set of scalar multiplets is denoted by $N^A$ which we decompose as $N^A=\{M^I,t_{\Lambda}\}$. The distinction is that the $t_{\Lambda}$ will be dualised to vector multiplets so that the $A^{\Lambda}$ become proper propagating fields. This is implemented by assuming constant shift gauging vectors and a constant embedding tensor $\Theta$ satisfying
\be
\tilde{X}^{\Lambda \Sigma} = -2i\delta^{\Lambda \Sigma} \;\;,\; \tilde{X}^{\Lambda I} = 0 \;\;,\; \Theta_{IJ}=0.
\ee 
This shift symmetry allows for the dualisation to vectors (which is not possible for general gauging vectors). If we assume that the submatrix $K_{(t)}=(K_{t_{\Lambda}t_{\Sigma}})$ is separately invertible with inverse $K_{(t)}^{t_\Lambda t_\Sigma}$ we can rewrite \eqref{eomA} as
\be
\label{eomAI}
\Theta_{\Lambda I}A^I = -\Theta_{\Lambda \Sigma}A^{\Sigma}+\tfrac12 d (\I\ t_{\Lambda})+\tfrac12 K_{(t)}^{t_\Lambda t_\Sigma}\I\  [K_{t_{\Sigma}M^I}\nabla M^I]-\tfrac{1}{8} K_{(t)}^{t_\Lambda t_\Sigma}*F^{\Sigma}.
\ee
We can use this to integrate out the non-propagating vectors $A^I$ by inserting \eqref{eomAI} into the action \eqref{kinetic_lin_gen_1app}.
The resulting action is given by $S =\int -\tfrac{1}{2}R_3 *1 - (V_\cT + V_F) * 1 + \cL_{1}$, where
\bea
\label{kinetic_lin_gen_dual}
\cL_{1} &=& -K_{M^I\bar{M}^{\bar J}}\nabla M^I\wedge * \nabla \bar{M}^{\bar{J}} -2\R \ (K_{t_{\Lambda}M^J}\nabla M^J)\wedge* d\R \ t_{\Lambda}  \nn \\&&
- K_{t_{\Lambda}t_{\Sigma}} d\R \ t_{\Lambda}\wedge* d\R \ t_{\Sigma} + K_{(t)}^{t_\Lambda t_\Sigma} \I \ [K_{t_{\Lambda}M^I}\nabla M^I]\wedge * \I \ [K_{t_{\Sigma}M^J}\nabla M^J] \nn \\&&
-\tfrac{1}{16}K_{(t)}^{t_\Lambda t_\Sigma}F^\Lambda \wedge * F^\Sigma +\tfrac{1}{2} \Theta_{\Lambda\Sigma}A^{\Lambda}\wedge F^\Sigma - \frac12 K_{(t)}^{t_\Lambda t_\Sigma} F^\Lambda \wedge \I \ [K_{t_{\Sigma}M^I}\nabla M^I].
\eea
We see that the imaginary parts of the $t_\Lambda$ have been dualized into the propagating vectors $A^\Lambda$. The real superpartners $\xi^\Lambda$ are obtained from $\R \ t_\Lambda$ by applying a Legendre transformation
\bea
K\left(t,M\right) &=& \tilde K\left(\xi,M\right) - \tfrac12 \left(t_{\Lambda}+\bar{t}_{\Lambda}\right) \xi^{\Lambda} \;, \label{legtran1}\\
\R \; t_{\Lambda} &=& \tilde K_{\xi^{\Lambda}} \label{legtran2} \;.
\eea
Note that due to the shift symmetry the K\"ahler potential $K\left(t,M\right)$ actually only depends on the $\R\; t_{\Lambda}$.

Following \cite{Grimm:2004uq} we differentiate (\ref{legtran1}) and (\ref{legtran2}) with respect to $t_{\Lambda}$ and $M^I$ to determine the useful relations
\bea
\frac{\partial \xi^{\Lambda}}{\partial t_{\Sigma}} &=& \tfrac12 \tilde{K}^{\xi^{\Lambda}\xi^{\Sigma}} \;\;,\; \frac{\partial \xi^{\Sigma}}{\partial M^{I}} = -\tilde{K}^{\xi^{\Lambda}\xi^{\Sigma}}\tilde{K}_{\xi^{\Lambda}M^I}\;, \nn \\ 
d\R\ t_\Lambda &=& \tilde{K}_{\xi^\Lambda \xi^\Sigma} d\xi^\Sigma + 2\R[\tilde{K}_{\xi^\Lambda M^I} d M^I] \;,\nn \\
K_{t_{\Lambda}} &=& -\tfrac12 \xi^{\Lambda} \;\;,\; K_{M^{I}} = \tilde{K}_{M^I}\;.
\eea
Differentiating again we deduce
\bea
&K_{t_\Lambda\bar{t}_\Sigma} = -\frac14 \tilde K^{\xi^{\Lambda}\xi^{\Sigma}} \;,\;\; & K^{t_\Lambda\bar{t}_\Sigma} = -4\tilde K_{\xi^{\Lambda}\xi^{\Sigma}} + 4 \tilde K_{\xi^{\Lambda}M^I} \tilde K^{M^I \bar{M}^{\bar{J}}} \tilde K_{\bar{M}^{\bar{J}}\xi^{\Sigma}} \;, \nn \\
&K_{t_\Lambda \bar{M}^{\bar{I}}} = \frac12 \tilde K^{\xi^{\Lambda}\xi^{\Sigma}} \tilde{K}_{\xi^{\Sigma}\bar{M}^{\bar{I}}} \;,\;\; & K^{t_\Lambda \bar{M}^{\bar{I}}} = 2 \tilde K^{\bar{M}^{\bar{I}}M^J} \tilde K_{M^J\xi^{\Lambda}} \;, \nn \\
&K_{M^I\bar{M}^{\bar{J}}} = \tilde K_{M^I\bar{M}^{\bar{J}}} - \tilde{K}_{M^I\xi^{\Lambda}} \tilde{K}^{\xi^{\Lambda}\xi^{\Sigma}} \tilde{K}_{\xi^{\Sigma}\bar{M}^{\bar{J}}} \;,\;\; & K^{M^I\bar{M}^{\bar{J}}} = \tilde{K}^{M^I\bar{M}^{\bar{J}}} \;. \label{transmet}
\eea
With the help of these relations we may rewrite
\bea
\label{kinetic_lin_gen_dual2}
\cL_{1} &=& -\tilde{K}_{M^I \bar{M}^{\bar{J}}}\nabla M^I\wedge * \nabla \bar{M}^{\bar{J}} + \tfrac{1}{4}\tilde{K}_{\xi^\Lambda \xi^\Sigma} d\xi^\Lambda\wedge * d\xi^\Sigma + \frac{1}{4}\tilde{K}_{\xi^\Lambda \xi^\Sigma}F^\Lambda \wedge * F^\Sigma  \nn \\&&
+\tfrac{1}{2}\Theta_{\Lambda\Sigma}A^\Lambda \wedge F^\Sigma + F^\Lambda \wedge \I[\tilde{K}_{\xi^\Lambda M^I}\nabla M^I] \nn \\&&
- d\xi^\Lambda \wedge* \R[\tilde{K}_{\xi^\Lambda M^I}\tilde{X}^{IJ}\Theta_{J\Sigma}A^\Sigma] \nn \\&&
-\tilde{K}^{\xi^\Lambda \xi^\Sigma} \R[\tilde{K}_{\xi^\Lambda M^I}\tilde{X}^{IJ}\Theta_{J\Sigma}A^\Sigma] \wedge * \R[\tilde{K}_{\xi^\Sigma M^{I'}}\tilde{X}^{I'J'}\Theta_{J'\Sigma}A^\Sigma].
\eea

Let us now focus on the case where $\tilde{X}^{IJ}$ is real. Then \eqref{DMThetaapp} describes the gauging of a shift symmetry of the real part of $M^I$, so that the K\"ahler potential must depend only on the imaginary part of $M^I$. This implies that $\tilde{K}_{\xi^\Lambda M^I}$ is purely imaginary, so that the last two lines of \eqref{kinetic_lin_gen_dual2} vanish.\\

We now aim to determine the transformed scalar potential. We return to the F-term piece $V_F$ later but first consider the `D-term' piece $V_\cT$ which is given in terms of the gauging
\bea
  i \partial_{A} \cP^{B} &=& K_{A\overline{C}} \tilde{X}^{\overline{C}B}\;, \nn \\ 
  \cT &=& - \tfrac12 \cP^{A} \Theta_{AB} \cP^{B}\;. \label{dtermgen}
\eea
To do this we also need to transform the derivatives
\be
\cT_{T_{\Lambda}} \to \tfrac12 \tilde K^{\xi^{\Sigma}\xi^{\Lambda}} \cT_{\xi^{\Sigma}} \;,\;\; \cT_{M^I} \to \cT_{M^I} - \tilde K^{\xi^{\Sigma}\xi^{\Lambda}} \tilde{K}_{\xi^{\Lambda}M^I} \cT_{\xi^{\Sigma}} \;.
\ee
This gives the resulting scalar potential
\beq
   V_\cT = \tilde{K}^{M^I \bar{M}^{\bar{J}}} \partial_{M^I} \cT \partial_{\bar{M}^{\bar{J}}} \cT - \tilde K^{\xi^{\Lambda}\xi^{\Sigma}} \partial_{\xi^{\Lambda}} \cT \partial_{\xi^{\Sigma}} \cT - \cT^2 \;.
\eeq 

We should also rewrite $\cT$. Since we already know the gauging vectors for the $t_{\Lambda}$ which we dualised we can write
\be
\cP^{\Lambda} = - \xi^{\Lambda} \;.
\ee
For the remaining chiral fields we need to transform the general expressions (\ref{dtermgen}). Taking the derivative index to be $t_{\Lambda}$ and $M^I$ gives the expressions
\be
i\partial_{t_{\Lambda}} \cP^{I} = K_{t_{\Lambda}\bar{M}^{\bar{J}}} \tilde{X}^{\bar{J}I} \;\;,
\qquad \quad i\partial_{M^K} \cP^{I} = K_{M^K\bar{M}^{\bar{J}}}\tilde{X}^{\bar{J}I} \;.
\ee
Transforming these gives
\bea
i\partial_{\xi^{\Lambda}} \cP^{I} &=& \tilde{K}_{\xi^{\Lambda}\bar{M}^{\bar{J}}} \tilde{X}^{\bar{J}I} \;, \label{trand1} \\ \nn
i\partial_{M^K}\cP^{I} - i\tilde{K}^{\xi^{\Lambda}\xi^{\Sigma}}\tilde{K}_{\xi^{\Sigma}M^K} \partial_{\xi^{\Sigma}} \cP^{I} &=&
 \left(\tilde{K}_{M^K\bar{M}^{\bar{J}}} - \tilde{K}_{M^K\xi^{\Lambda}}\tilde{K}^{\xi^{\Lambda}\xi^{\Sigma}}\tilde{K}_{\xi^{\Sigma}\bar{M}^{\bar{J}}} \right) \tilde{X}^{\bar{J}I} \;. 
\eea
Using the first equation of (\ref{trand1}) in the second gives the expression
\be
i\partial_{M^K}\cP^{I} = \tilde{K}_{M^K\bar{M}^{\bar{J}}} \tilde{X}^{\bar{J}I}  \;.
\ee
This is very similar to the D-term equation in four dimensions. To make a closer match we define 
\be
   D_{\Sigma} = \Theta_{\Sigma I} \cP^{I} \;\;,\; X^{\bar{J}}_{\Sigma} =  \Theta_{\Sigma I} \tilde{X}^{\bar{J}I} \;,
\ee
which gives
\be
i\partial_{M^K} D_{\Sigma} = \tilde{K}_{M^K\bar{M}^{\bar{J}}} X^{\bar{J}}_{\Sigma}  \;.
\ee
This is of the same form as the usual four-dimensional D-term equation. Finally we can write $\cT$ in terms of the transformed quantities
\be
\cT = -\tfrac12 \xi^{\Sigma} \Theta_{\Sigma\Lambda} \xi^{\Lambda} +  D_{\Sigma} \xi^{\Sigma}\;.
\ee
This completes the dualisation of the D-term piece $V_{\cT}$. 

The F-term piece $V_F$ of the scalar potential will not play a role in our calculations but for completeness we also transform it here. To do so we note that 
\bea
D_{T_{\Lambda}} W &\to& -\frac12 \xi^{\Lambda} W \;, \nn \\
D_{M^I} W &\to& \left[ \partial_{M^I} W + \left(\partial_{M^I}K\right)W \right] + \tilde{K}_{\xi^{\Lambda}M^I}\xi^{\Lambda} W \equiv D_{M^I} W + \tilde{K}_{\xi^{\Lambda}M^I}\xi^{\Lambda} W \;.   
\eea
Where we henceforth denote the part in the square brackets as $D_{M^I} W$ noting that in it the K\"ahler potential rather than the kinetic potential appears. With these transformations and using (\ref{transmet}) we find
\be
V_F = e^K \left[ \tilde{K}^{M^I\bar{M}^{\bar{J}}} D_{M^I} W D_{\bar{M}^{\bar{J}}} \overline{W} - \big(4+\tilde{K}_{\xi^{\Lambda}\xi^{\Sigma}}\xi^{\Lambda}\xi^{\Sigma} \big) |W|^2 \right] \;. 
\ee

\section{Breaking $D_{n} \to A_{n-1}\times U(1)$}
\label{Dnbreaking}

In section \ref{sec:u1geometry} we considered deforming an $A_{n+m-1}$ singularity to an $A_{n-1}\times A_{m-1}$ singularity in order to identify the non-local realisation of the $U(1)$ associated to two stacks of branes. In this appendix we consider deforming a $D_n$ singularity to an $A_{n-1}$ singularity which will allow us to identify the non-local realisation of the $U(1)$ associated to a brane and image pair. 
The deformation of a $D_n$ singularity from the perspective of the cycle geometry has been discussed  in great detail in \cite{Braun:2008ua}.

The defining equation for a $D_n$ singularity is
\be
y^2 = -x^2 z + z^{n-1} \;.
\ee
We can resolve it to preferred versal form
\be
y^2 = -x^2 z + z^{n-1} + \sum_i^{n-1} \delta_{2i} z^{n-i-1} - 2 \gamma_n x \;,
\ee
with $\delta_{2i}$ being the elementary symmetric polynomials of degree $i$ in the $t_i^2$ and $\gamma_n=\prod_i^n t_i$. The deformation parameters $t_i$ are similar to those of an $A_{n-1}$ singularity except that now there is no traclessness constraint (\ref{tracelessness}). A compact way to write this is 
\be
y^2 = -x^2 z + \frac{1}{z}\left( g_{D_n} - f_{D_n}^2 \right)  - 2 x f_{D_n}  \;,
\ee
with the explicit form being
\be
y^2 = -x^2 z + \frac{1}{z}\left[ \prod_i^n\left(z+t_i^2\right)-\prod_i^n t_i^2 \right]  - 2 x \prod_i^n t_i  \;.
\ee
Identifying the 2-cycles is more complicated than the $A_{n-1}$ case. We can rewrite the equation as
\be
\left(y'\right)^2 + \left(x'\right)^2 = \prod_i^n\left(z+t_i^2\right) \;,
\ee
with $y'=\sqrt{z}y$ and $x'=xz+f$. Then we have a collapsing $S^1$, given by restricting to the real $(x,y)$-planes, at $z=-t_i^2$. However tracking the fibration of this $S^1$ over a real line in the $z$-plane is more complicated since the coordinate $y'$ has branch cuts. For the cycles where the path through $\tilde{z}$ connecting the $t_i$ does not go through the branch cut the cycles are simply constructed as in the $A_{n-1}$ case. For the cases where the path crosses the branch cut we can go to the double cover $\tilde{z}=\sqrt{z}$ giving 
\be
\left(\tilde{y}'\right)^2 + \left(\tilde{x}'\right)^2 = \prod_i^n\left(\tilde{z}^2+t_i^2\right) \;,
\ee
with $\tilde{y}'=\tilde{z}y$ and $\tilde{x}'=x\tilde{z}^2+f$. Now we have doubled the degeneration points since they occur at $\tilde{z} = \pm i t_i$. This is now two copies of a resolved $A_{n-1}$ singularity but with also the possibilities of constructing cycles between the two. Indeed we can identify the ${\mathbb P}^1$ by restricting to the real subspace of $(\tilde{x}',\tilde{y}')$ and the imaginary one of $\tilde{z}'$ while considering a path between say $it_1$ and $-it_1$ so that
\be
\left(\R\; \tilde{y}'\right)^2 + \left(\R\; \tilde{x}'\right)^2 + \left(\I\; \tilde{z}\right)^2 = t_1^2\;, \label{p1or}
\ee
giving the ${\mathbb P}^1$. In the IIB limit these are string stretching between the brane and its image.

The simple roots are constructed as fibrations over the lines $v_i=t_i - t_{i+1}$ with $i=1,...,n-1$ and $v_n = t_{n-1} + t_n$. In the single cover we connect $t_i$ points as before but now can also connect them by circling around $z=0$ or not thereby giving the possibility of the $v_n$ root. In the double cover the sign choice corresponds to connecting $+it_i$ with $\mp it_j$.

To see the origin of the $U(1)$ we should consider the resolving the singularity $D_n \rightarrow A_{n-1}\times U(1)$.  
This breaking is done by taking $\left(t_1,...,t_n\right) = \left(t,...,t\right)$. This gives a $D_n$ singularity at $t=0$ and also gives an $A_{n-1}$ singularity for $t \neq 0$ located at $z=-t^2$ \cite{Morrison:1996na}. Therefore this is describing the separation of brane stacks from the orientifold. The simple roots for the $D_n$ singularity are $v_i = 0$ for $i\neq n$. This, in the mirror picture, identifies the diagonal $U(1)$ as the resolution combination $\tilde{v}_n$.



\end{document}